\documentclass[aps,prc,reprint,superscriptaddress,amsmath,amssymb,nofootinbib]{revtex4-2}

\usepackage{setspace}
\usepackage{graphicx}
\usepackage{bm}
\usepackage[utf8]{inputenc}
\usepackage[colorlinks=true,allcolors=blue]{hyperref}
\usepackage{tabularx}
\usepackage{aasmacros}
\begin{document}

\title{First strength measurements of low-energy resonances in the $^{45}$Sc($p,\gamma$)$^{46}$Ti reaction and its astrophysical implications}

\author{R.~S.~Sidhu}
\email{ragan.sidhu@surrey.ac.uk}
\affiliation{School of Mathematics and Physics, University of Surrey, Guildford, GU2 7XH, United Kingdom}

\author{R.~J.~deBoer}
\affiliation{Department of Physics and Astronomy, University of Notre Dame, Notre Dame, Indiana 46556, USA}

\author{J.~Görres}
\affiliation{Department of Physics and Astronomy, University of Notre Dame, Notre Dame, Indiana 46556, USA}

\author{M.~Wiescher}
\affiliation{Department of Physics and Astronomy, University of Notre Dame, Notre Dame, Indiana 46556, USA}

\author{W.~N.~Catford}
\affiliation{School of Mathematics and Physics, University of Surrey, Guildford, GU2 7XH, United Kingdom}

\author{C.~Dembski}
\affiliation{Department of Physics and Astronomy, University of Notre Dame, Notre Dame, Indiana 46556, USA}

\author{C.~R.~Jones}
\affiliation{Department of Physics and Astronomy, University of Notre Dame, Notre Dame, Indiana 46556, USA}

\author{Ó.~E.~López-López}
\affiliation{Instituto de Física, Universidad Nacional Autónoma de México, 04510, Ciudad de México, Mexico}

\author{G.~Lotay}
\affiliation{School of Mathematics and Physics, University of Surrey, Guildford, GU2 7XH, United Kingdom}

\author{K.~Manukyan}
\affiliation{Department of Physics and Astronomy, University of Notre Dame, Notre Dame, Indiana 46556, USA}

\author{M.~Matney}
\affiliation{Department of Physics and Astronomy, University of Notre Dame, Notre Dame, Indiana 46556, USA}

\author{J.~O'Neill}
\affiliation{School of Physics and Astronomy, University of Birmingham, Birmingham, B15 2TT, United Kingdom}

\author{J.~Rufino}
\affiliation{Department of Physics and Astronomy, University of Notre Dame, Notre Dame, Indiana 46556, USA}

\author{A.~T.~Sanchez}
\affiliation{Department of Physics and Astronomy, University of Notre Dame, Notre Dame, Indiana 46556, USA}

\author{M.~Williams}
\affiliation{School of Mathematics and Physics, University of Surrey, Guildford, GU2 7XH, United Kingdom}

\author{L.~Zimmer}
\affiliation{Department of Physics and Astronomy, University of Notre Dame, Notre Dame, Indiana 46556, USA}

\date{\today}

\begin{abstract}
The $^{45}$Sc($p,\gamma$)$^{46}$Ti reaction plays an important role in both hydrostatic and explosive nucleosynthesis. 
During hydrostatic oxygen and silicon burning in massive stars, including dynamic processes such as convective carbon-oxygen (C-O) shell mergers, it acts as a bottleneck that regulates the reaction flow in the Sc-Ti mass region. 
In oxygen-neon (ONe) novae, the ejecta indicate the production of elements up to the Fe group, as observed in V1974 Cygni, and the $^{45}$Sc($p,\gamma$)$^{46}$Ti reaction links nuclei above Ca to the Fe region. In core-collapse supernovae (CCSNe), it influences the production of the radioactive isotope $^{47}$Sc.
At present, the thermonuclear reaction rate is based on statistical model calculations.
In this work, we report on measurements of the $^{45}$Sc($p,\gamma$)$^{46}$Ti reaction performed using the 5~MV accelerator at the Nuclear Science Laboratory, University of Notre Dame, over a laboratory proton energy range of 526--1275~keV. We present the first direct measurements of the resonance strengths for six 
resonances at $E^{\rm lab}_{\rm R} = 917.4$, 1026.0, 1031.1, 1049.7, 1059.4 and 1257.5~keV, together with integrated resonance strengths at $E^{\rm lab} = 555$, 630, 700, 761 and 871~keV. 
For resonance energies below the lowest directly measured energy, available transfer-reaction data were used to estimate the resonance strengths. 
An updated thermonuclear reaction rate is derived, and its astrophysical implications are discussed.

\end{abstract}


\maketitle

\section{Introduction}

Nuclear reaction rates for capture processes on isotopes in the \textit{fp} shell remain poorly constrained and carry considerable uncertainties. These reactions are of particular importance for nucleosynthesis pathways that connect closed shell nuclei such as $^{28}$Si and $^{40}$Ca to the Fe-Ni mass region through proton capture reactions near the line of stability, which are, compared to alpha capture reactions, less hindered by the Coulomb barrier. Currently, the reaction rates are derived primarily from statistical model calculations using standard reaction codes such as \texttt{TALYS}~\cite{Koning2023} and \texttt{NON-SMOKER}~\cite{RAUSCHER2001}. These calculations rely on the assumption of sufficiently high-level densities in both the compound and final nuclei. However, this assumption may not be valid for nuclei near the closed \textit{Z}~=~20 shell, and therefore requires experimental verification over the temperature range relevant to astrophysical sites where these processes occur and play a crucial role~\cite{PhysRevC.56.1613}.

There are three astrophysical sites that provide sufficiently high hydrogen abundance and temperature conditions for such reactions to occur. The first, and perhaps the most intriguing, is during carbon and oxygen (C-O) shell mergers during the late stellar evolution of massive stars~\cite{meakin2006active,roberti2025occurrence}. 
This convective-reactive event occurs when the convective mixing timescale becomes comparable to the nuclear burning timescale, generating a mixed convective region in which carbon-rich material from the outer carbon shell is rapidly ingested into the hot underlying oxygen-burning layer.
At these oxygen-burning temperatures, the ingested $^{12}$C and ambient $^{16}$O undergoes fusion through reactions such as $^{12}$C+$^{12}$C, $^{12}$C+$^{16}$O, and $^{16}$O+$^{16}$O, along with $^{20}$Ne photodisintegration, suddenly releasing a strong flux of light particles, primarily protons and $\alpha$ particles~\cite{roberti2025occurrence}. These released light particles are subsequently captured by oxygen-burning ashes, driving the nucleosynthesis flow that produces intermediate odd-$Z$ nuclei such as Cl, K, and Sc~\cite{2018MNRAS.480..538R, 2020MNRAS.491..972A}. In these nucleosynthesis models, $^{34}\text{S}$ is a primary seed nucleus~\cite{roberti2025occurrence}: proton capture forms $^{35}\text{Cl}$, while successive $\alpha$ and proton captures produce $^{39}\text{K}$ via $^{34}\text{S}(\alpha,\gamma)^{38}\text{Ar}(p,\gamma)^{39}\text{K}$. In parallel, $\alpha$ capture on $^{38}\text{Ar}$ forms $^{42}\text{Ca}$, which is converted to $^{45}\text{Sc}$ through $^{42}\text{Ca}(\alpha,p)^{45}\text{Sc}$~\cite{2018MNRAS.480..538R,2020MNRAS.491..972A,roberti2025occurrence}.
As the reaction flow proceeds to higher masses, the subsequent $^{45}$Sc($p,\gamma$)$^{46}$Ti reaction is considered a bottleneck for the pathway towards higher masses, during both the oxygen and silicon burning stages in massive stars~\cite{1978ApJ...223..697S,2014JPSCP...1a3054O,woosley1973explosive}.
More recently, studies have further suggested that radioactive nuclei along the $\alpha$ chain, such as $^{44}$Ti ($t_{1/2} = 59.1(3)$~y), may also be synthesized during C-O shell mergers~\cite{Boccioli:2026kfw, issa2026}.

The production of such radioactive nuclei is also closely connected to a second important astrophysical environment, namely core-collapse supernovae (CCSNe), where explosive silicon burning gives rise to a range of radioactive isotopes whose abundances are highly sensitive to the underlying nuclear reaction rates. Sensitivity studies of nucleosynthesis in CCSNe identify the $^{45}$Sc($p,\gamma$)$^{46}$Ti reaction as one of the reactions that significantly affects the production of the radioactive isotope $^{47}$Sc ($t_{1/2} = 3.3491(6)$~d), which is synthesized in the innermost ejecta during the explosion~\cite{hermansen2020reaction}. Radioactive $^{47}$Sc decays via $\beta^-$ emission to $^{47}$Ti, producing a characteristic $\gamma$-ray at $E_\gamma$~=~159.4~keV. Observation of this $\gamma$-ray line by future space-based $\gamma$-ray telescopes~\cite{timmes2019catching} would provide a direct probe of explosive nucleosynthesis, offering constraints on both nucleosynthesis models and the physics of CCSNe.

The third environment of interest is the  
oxygen-neon (ONe) nova. The lack of experimental reaction rates above the closed shell nucleus $^{40}$Ca represents an uncertainty with respect to the endpoint of the nucleosynthesis pattern. ONe novae have been identified as thermonuclear runaways on the surfaces of O-Ne-Mg enriched white dwarfs, which develop after the carbon-burning phase during stellar evolution. The reaction path above the seed nucleus $^{20}$Ne is characterized by the hot NeNa and MgAl cycles, which produce long-lived radioactive nuclei $^{22}$Na and $^{26}$Al, respectively~\cite{1998ApJ...494..680J, 2014MNRAS.442.2058D, 1989ApJ...346.1041W}. Extensive measurements of the proton capture reactions in the higher mass range have demonstrated that the reaction path above $^{28}$Si and $^{32}$S follows the line of stability towards $^{40}$Ca~\cite{1995PhRvC..52.1681R, 1994NuPhA.571..132I, 1993NuPhA.559...83I, 1992PhRvC..45.2989I, 1991NuPhA.533..153I, 1992NuPhA.539...97I, 2023PhRvC.107f5806S}. At temperatures typical of ONe novae, 0.2~$< T <$~0.4~GK~\cite{downen2013nuclear}, the reaction rates are too slow to compete with the $\beta^+$ decays of nuclei along the path, and back-processing reactions $^{31}$P($p,\alpha$)$^{28}$Si, $^{35}$Cl($p,\alpha$)$^{32}$S, and $^{39}$K($p,\alpha$)$^{36}$Ar are hindered by the reduced penetrability of $\alpha$ particles through the Coulomb barrier of their respective high-\textit{Z} compound nuclei~\cite{1995PhRvC..52.1681R}. The endpoint of the reaction path has been linked with proton capture on the doubly magic $^{40}$Ca nucleus, but a recent measurement of the  $^{40}$Ca($p,\gamma$)$^{41}$Sc reaction suggests a reasonably strong reaction rate, indicating that a small amount of material can be processed toward higher masses~\cite{sidhu2025strength}. While it was shown that ONe novae ejecta are dominated by medium mass abundances in the Al to Si range, observations such as those of Cygni V1974 indicate the presence of Fe material, which might be explained by a reaction path bridging into higher mass ranges~\cite{1997AstL...23..713A, 2002ApJ...577..940S}. However, these interpretations still carry significant uncertainties, as the physical processes in the ejecta environments are not well understood.

For nucleosynthesis pathways beyond $^{40}$Ca, $^{45}$Sc is the only stable scandium isotope, while the more neutron-deficient Sc isotopes are generally short-lived, with the exception of $^{44}$Sc, which has a half-life of $t_{1/2}=4.0420(24)$~h. Consequently, proton capture on either $^{44}$Sc or $^{45}$Sc may provide a pathway towards the synthesis of heavier nuclei. 
Furthermore, $^{45}$Sc($p,\gamma$)$^{46}$Ti has recently been identified as a key reaction governing the abundance uncertainty of $^{45}$Sc in CO and ONe nova models, thereby directly affecting nucleosynthesis predictions for the symbiotic recurrent nova T Coronae Borealis (T~CrB)~\cite{wallace2025elemental}.
For this reason, a measurement of the $^{45}$Sc($p,\gamma$)$^{46}$Ti reaction has been performed to probe the contributions of low-energy resonances, which have never been investigated~\cite{1994PAN....57..727N, 1978ApJ...223..697S}.
\begin{figure}
   \centering
    \includegraphics[width=0.48\textwidth]{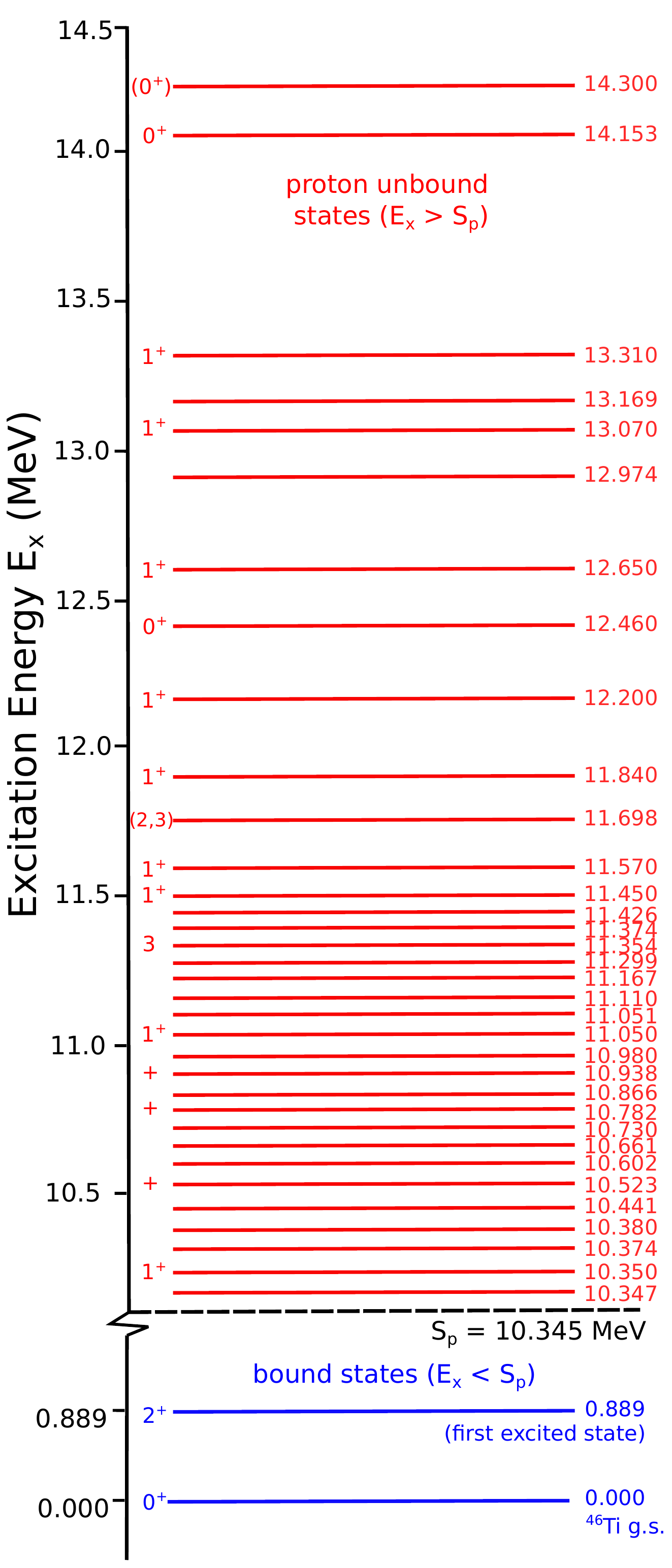}
  \caption{Known energy-level scheme of the $^{46}$Ti compound nucleus~\cite{WU20001}. The $^{45}$Sc($p,\gamma$)$^{46}$Ti reaction is exothermic, with a $Q$-value of 10344.8(7)~keV, where the known proton-unbound states are indicated by the red lines. In the present work, direct measurements were performed for resonances in the proton energy range 526--1275~keV, corresponding to excitation energies of $E_x = 10.870$--11.620~MeV in $^{46}$Ti. For all resonances, the secondary $E2$ transition from the first excited state to the ground state at $E_\gamma = 889.277(3)$ keV (hereafter referred to as the 889 keV transition) was used to determine the resonance strengths.}
  \label{fig:Fig-Level_Scheme}
\end{figure} Recent studies have emphasized that the total reaction rate should include contributions from proton capture on isomeric states~\cite{1978ApJ...223..697S, 2024EPJST.233.1075M}. The present study focuses only on proton capture on the ground state.

\section{The compound structure of $^{46}$Ti}
\label{sec:Literature_Review}

The cross section for the $^{45}$Sc($p,\gamma$)$^{46}$Ti radiative capture reaction 
has been studied previously in three experiments~\cite{Dubois1963,1978NuPhA.312..140S,1987NCimA..97..211M}. Complementary proton-transfer studies have been performed~\cite{broman1968level,broman1971the45sc} to investigate the level structure above the proton threshold. 
The currently known level structure of the compound nucleus $^{46}$Ti is shown in Fig.~\ref{fig:Fig-Level_Scheme}. A brief overview of previous experimental studies relevant to its nuclear structure is presented below.

\subsection{Direct Measurements}
The first direct study, conducted by~\citet{Dubois1963} in 1963, employed a Van de Graaff accelerator at Chalmers University of Technology to produce proton beams with energies between 900 and 1400~keV, which were directed onto a thin $^{45}$Sc target mounted on a 0.2~mm thick Ta backing. For detection, two NaI(Tl) crystals of dimensions 5~$\times$~4 inches, each coupled to an E.M.I.~9583 photomultiplier tube, were positioned opposite one another at 90$^{\circ}$ to the beam direction.
The relative excitation curve of the $^{45}$Sc($p,\gamma$)$^{46}$Ti reaction was measured twice across the 900--1400~keV range and was found to be reproducible.  
In total, 134 resonances were identified. 
Capture $\gamma$-rays were studied at ten of these resonances, including those at 917.4, 1026.0, 1031.1, 1049.7, 1059.4 and 1257.5~keV, which are reported in this study. 
A level scheme for $^{46}$Ti was proposed, in which all populated levels were observed to decay to the first excited state, which subsequently de-excited to the ground state via emission of an $E2$ $\gamma$-ray at 889~keV.
Despite the detailed excitation curves and decay scheme presented in that work, no efficiency corrections were applied, and no absolute resonance strengths were reported. As the reported relative $\gamma$-ray intensities were based on transitions above 1.5~MeV, whereas the present work uses the 889~keV $\gamma$ ray, this limits the direct quantitative use of their relative yield measurements in the current analysis. A more detailed discussion of comparison with the study of~\citet{Dubois1963} is provided in Sec.~\ref{subsection:Resonance scan} and subsequent sections. 

The second direct study, conducted by~\citet{1978NuPhA.312..140S}, investigated the reaction over a wide proton energy range of 0.7--4.2~MeV. The target, with a thickness of 50~$\mu$g/cm$^2$, was prepared by evaporating 99.99\% pure scandium from a tungsten wire basket onto a 0.25~mm thick tantalum backing. Proton beams were provided by the Melbourne 5~MV Pelletron accelerator. The targets were positioned at an angle of 55$^{\circ}$ to the incident beam, with a Ge(Li) detector placed 11~mm from the target at the same angle. To reduce background, the detector was shielded with a 3~mm thick lead sheet to suppress the intense 136~keV $\gamma$-ray arising from inelastic scattering and Coulomb excitation of the tantalum backing. 
In that work, observation of the 889~keV $\gamma$-ray peak provided a reliable measure of the $\gamma$-ray yield for proton energies up to 2.5~MeV. At higher energies, however, the presence of the 890.9~keV $\gamma$-ray, originating from the population and subsequent de-excitation of the 1.433~MeV state in scandium via the ($p,p^{\prime}\gamma$) reaction, rendered extraction of the 889~keV yield impossible. Instead, the 1120.5~keV $\gamma$-ray, corresponding to the transition between the 2.009 and 0.889~MeV states in $^{46}$Ti, was used for yield determination. This study provided absolute measurements of the cross section and derived thermonuclear reaction rates, but, no resonance strengths were reported.

The third direct study was carried out by~\citet{1987NCimA..97..211M}. 
The experiment was performed using the proton beams from the 3~MeV Van~de~Graaff accelerator of the Atomic Energy Centre in Dhaka, Bangladesh. A target of nominal thickness, 20~$\mu$g/cm$^2$, was prepared by evaporating a scandium layer on Ta backing. Thicker targets, 100~$\mu$g/cm$^2$, were used for the detection of secondary and weaker primary $\gamma$ rays. 
The $\gamma$-rays were detected using a 30 cm$^2$ Ge(Li) detector placed on a movable base.
Spectra were recorded at 0$^{\circ}$, 30$^{\circ}$, 55$^{\circ}$, and 90$^{\circ}$ to measure the angular distributions of the $\gamma$-rays; alternatively, measurements were taken only at 55$^{\circ}$ and 90$^{\circ}$ relative to the beam axis to determine the $\gamma$-ray branching ratios and energies. The study measured the decay schemes of the resonances at 
$E^{\rm{lab}}_{\rm{R}}$ = 1026, 1047, and 1377~keV, and the angular distributions at 1026 and 1377~keV; however, no resonance strengths were reported.

\subsection{Transfer reaction studies}

For the transfer reaction measurements, we restrict our discussion to proton-transfer reactions.

The first particle transfer-based study of the level structure of $^{46}$Ti was reported by \citet{broman1968level}. The $^{45}$Sc($^{3}$He,d)$^{46}$Ti proton stripping the and $^{47}$Ti($^{3}$He,$\alpha$)$^{46}$Ti neutron pick-up reactions were studied using a 15~MeV $^{3}$He$^{2+}$ beam from the University of Pennsylvania tandem accelerator. The charged reaction products were momentum analyzed with a broad-range magnetic spectrograph and recorded in nuclear emulsions.
Deuteron angular distributions were measured between 7 and 60$^{\circ}$ for transitions populating 106 states up to an excitation energy of 11.4~MeV in $^{46}$Ti. Using distorted-wave Born approximation (DWBA) analysis, proton orbital angular momentum transfers and spectroscopic strengths were extracted for states up to 7.6~MeV.
From the ($^{3}$He,d) reaction, 15 proton-unbound levels were identified, as listed in Table~2 of Ref.~\cite{broman1968level}. 
Among these, the lowest seven proton-unbound states are relevant for the present reaction-rate evaluation and are discussed further in Sec.~\ref{sec:Transfer_reaction_data}. However, for all 15 proton-unbound states, neither proton orbital angular momentum transfers nor spectroscopic strengths were reported, and the excitation energies were reported with an uncertainty of 25~keV. 

A subsequent analysis by \citet{broman1971the45sc} revisited the same $^{45}$Sc($^{3}$He,d)$^{46}$Ti data~\cite{broman1968level}, extending the DWBA comparison to include stronger transitions to 29 states between 7.6 and 11.4~MeV excitation energy. In this later study, proton-transfer assignments and spectroscopic factors were extracted for three proton-unbound levels (see Table~1 of \citet{broman1971the45sc}). The quoted uncertainty in the cross sections, and hence in the spectroscopic factors, is $\pm$30\%, arising primarily from uncertainties in the choice of potential parameters, the target thickness determination, and background contributions to weak transitions.

The $^{45}$Sc($^{3}$He,d)$^{46}$Ti reaction was also studied by \citet{banu1982level} using the Van de Graaff accelerator and the multi-channel magnetic spectrograph at the Nuclear Physics Laboratory of University of Oxford. However, only proton-bound levels up to $E_x$ = 9.9~MeV in $^{46}$Ti were investigated, which are not relevant to the present study.

As evident from previous measurements and literature, no resonance strengths have yet been determined for the low-energy resonances in the $^{45}$Sc($p,\gamma$)$^{46}$Ti reaction, which motivated the present investigation.

\section{Experiment}
\label{sec:Experiment}

\subsection{Experimental setup}
\begin{figure}
   \centering
    \includegraphics[width=0.5\textwidth]{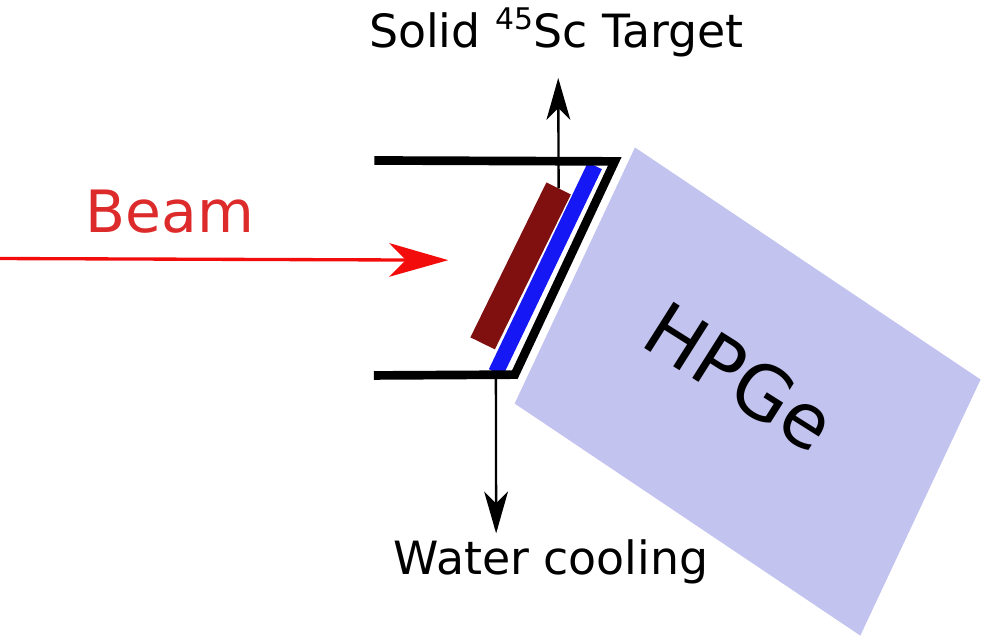}
  \caption{Experimental setup for the $^{45}$Sc($p,\gamma$)$^{46}$Ti reaction measurement in close geometry. A HPGe detector was placed at 45$^{\circ}$ relative to the beam direction, directly behind the target holder. The setup was surrounded by lead shielding to reduce background from nearby radiogenic radiation.}
  \label{fig:Fig-Setup}
\end{figure}

The experiment was performed using the St.~ANA 5~MV Pelletron accelerator at  
Nuclear Science Laboratory (NSL) at the University of Notre Dame~\cite{aprahamian2014nuclear}. The 5~MV accelerator delivered proton beams with an intensity limited to $<$~5~$\mu$A on a Sc target. The energy calibration of the accelerator was performed using a well-known resonance in the $^{27}$Al$(p,\gamma)^{28}$Si reaction at $E^{\rm{lab}}_{\rm{R}}$~=~992~keV~\cite{antilla1977use}. The beam was sent through a 90$^\circ$ analyzing magnet before being transported to the experimental target station, providing a beam energy uncertainty of $\pm$1~keV at 1~MeV. The beam was defocused and wobbled over an area of approximately 1~cm$^2$ on the target to mitigate target degradation and reduce the effects of target inhomogeneities. The target was mounted at 45$^\circ$ relative to the incoming beam as shown in Fig.~\ref{fig:Fig-Setup}.

A high purity germanium (HPGe) detector, a p-type coaxial EGC 100-260-R from Canberra~\cite{hpge}, was used to measure $\gamma$-rays from the $^{45}$Sc($p,\gamma$)$^{46}$Ti reaction. This HPGe detector was quoted to have an efficiency of 130\% relative to a 3~in.~$\times$~3~in. NaI detector, with an energy resolution of $\approx$ 0.26\% at 1332~keV. The detector was placed at 45$^\circ$ relative to the beam direction on a movable stand, directly behind the target as shown in Fig.~\ref{fig:Fig-Setup}. 
To attenuate low-energy $\gamma$ rays from the $^{45}$Sc($p,p'\gamma$)$^{45}$Sc reaction, the front face of the HPGe detector was covered with a 2.83~mm (3/32$^{\rm nd}$ in.) thick lead sheet. The electrically isolated target chamber acted as a Faraday cup, which was used for the integration of the beam current. To limit carbon buildup on the target, a cold trap was mounted that consisted of a copper tube, cooled to liquid nitrogen temperature, and placed inside the beam line. The copper tube extended to within a few millimeters of the target surface. The cold trap was electrically insulated and biased to $-$300~V to suppress secondary electrons. The target was water-cooled to reduce damage resulting from beam-induced power deposition.

\begin{figure*}
   \centering
    \includegraphics[width=1\textwidth]{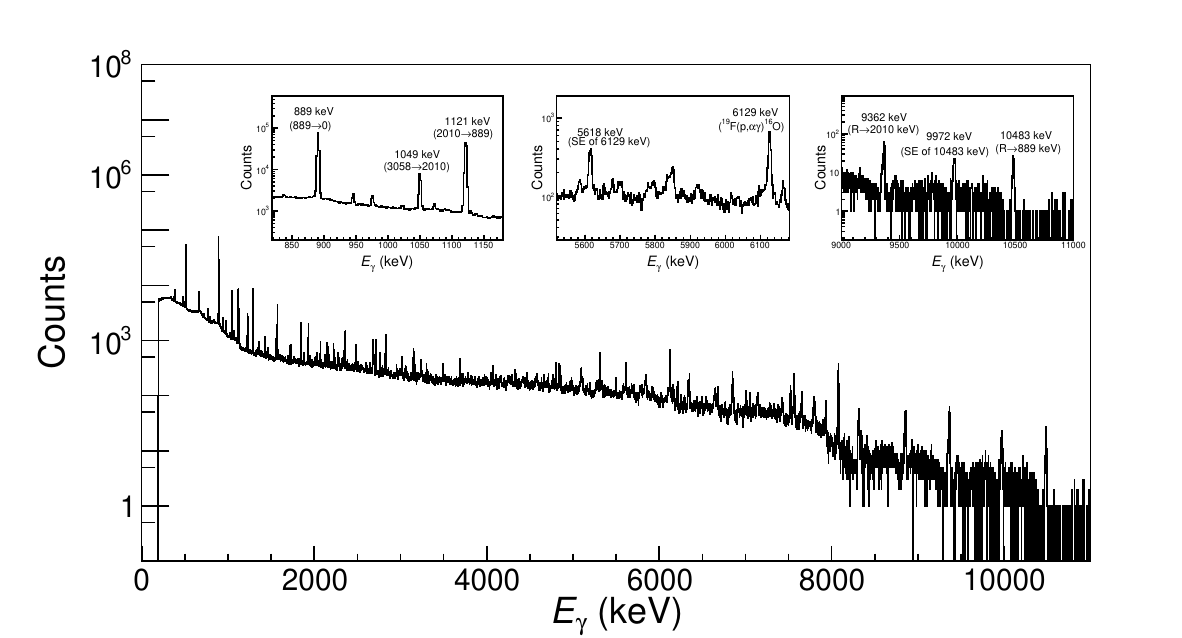}
  \caption{A $\gamma$-ray spectrum from a long run taken at far geometry for the $E^{\rm{lab}}_{\rm{R}} = 1049.7$~keV resonance. The insets show three selected $\gamma$-ray energy regions: the first shows the 889 and 1121~keV secondary $\gamma$ rays, the second shows $\gamma$ rays from $^{19}$F($p,\alpha\gamma$)$^{16}$O, and the third covers the high-energy region where primary transitions to the 889 and 2010~keV levels are visible.
  }
  \label{fig:Gamma_ray_1050keV}
\end{figure*}

\subsection{Target Preparation}
A fresh Sc target ($\approx$50 $\mu$g/cm$^2$)
was prepared using the vacuum evaporation technique. Since the target was mounted at $45^\circ$ (see Fig.~\ref{fig:Fig-Setup}), the effective thickness becomes $\approx$71 $\mu$g/cm$^2$. This corresponds to effective target thicknesses ($\Delta E$) of $\approx$ 9 to 11~keV over the proton beam energies of corresponding to the resonances at 917.4, 1031.1 and 1257.5 keV, respectively\footnote{This target thickness was chosen to satisfy the thick-target condition~\cite{1948RvMP...20..236F} ($\Gamma << \Delta E$; with $\Gamma < 0.1$~keV); the partial widths and their relation to the resonances strengths are discussed in Sec.~\ref{subsubsection:below917keV})  (see Eq.~\ref{eq:equation-4}).}.
A 0.5~mm thick Ta sheet was used as the target backing. It was sequentially cleaned with acetone, ethyl alcohol, and semiconductor-grade isopropyl alcohol, with ultrasonication applied at each step, followed by drying in an Ar flow. The sheet was then annealed under a vacuum of 1$\times$10$^{-6}$~Torr using direct current heating to remove surface contamination. Sc metal (natural enrichment) was subsequently evaporated at a pressure of 8$\times$10$^{-6}$~Torr. The distance between the Sc source boat and the tantalum sheet was set at 20~cm to ensure uniform deposition. Real-time monitoring of the deposition was performed using Au-coated quartz crystals, with thickness measurements recorded by a monitor. The evaporation rate was kept at 0.5~$\mu$g/min. After the evaporation process, the target was allowed to cool for 30 minutes, followed by introducing high-purity Ar gas into the evaporator. Finally, the targets were transferred to the beamline using an Ar-filled glove bag to prevent oxidation.

\subsection{Experimental Procedure}

The main aim of the experiment was to measure the strongest resonances in the proton beam energy range $E^{\rm{lab}}_{\rm{R}} = 900$--1270~keV, determine their strengths and thereby their reaction rates, and compare the results with statistical-model predictions. In this energy region, the strongest resonances identified by \citet{Dubois1963} are at 917.4, 1026.0, 1049.7 and 1257.5~keV\footnote{The 1031.1 and 1059.4~keV resonances were additionally measured during a scan of the 1026.0 and 1049.7~keV resonances (see Sec.~\ref{subsubsection:1026-scan}).}. The experiment was performed in the following steps.

\begin{figure*}
   \centering
    \includegraphics[width=0.85\textwidth]{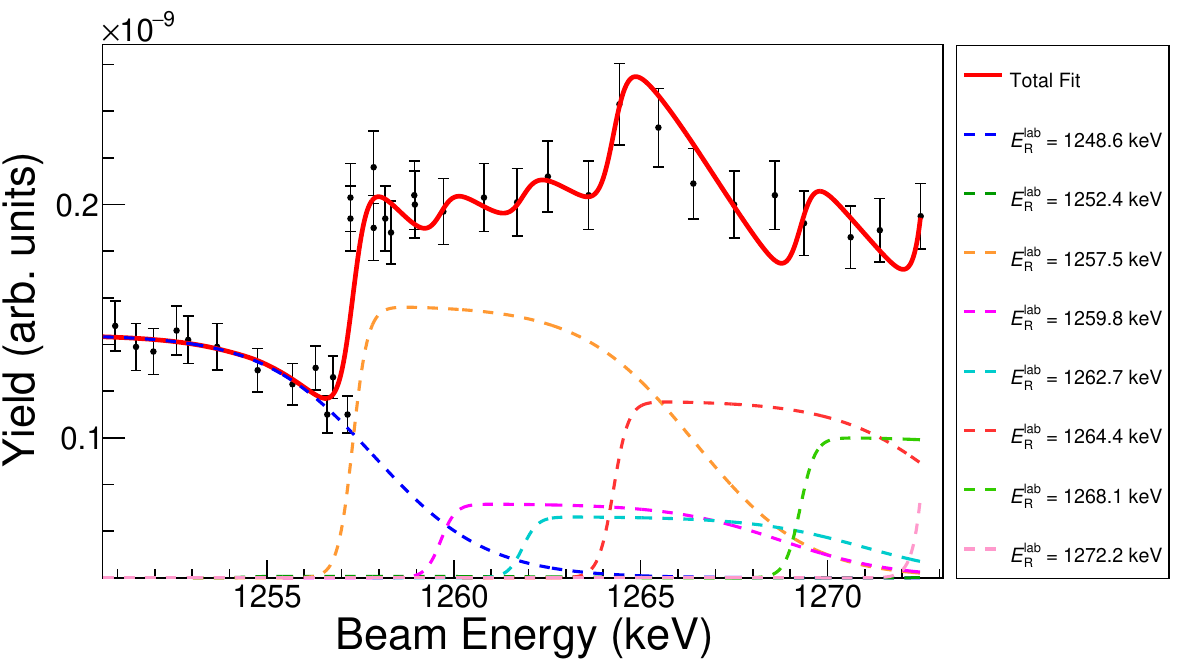}
  \caption{Resonance scan plot for the $E^{\rm{lab}}_{\rm{R}}$~=~1257.5~keV resonance using the 889~keV secondary $\gamma$-ray transition. The red curve shows the total fit, while the dotted lines indicate contributions from individual resonances. 
  The target thickness obtained using Eq.~(\ref{equation-1}) is $\Delta E^{\rm lab} = 9.1(5)$~keV.
  }
  \label{fig:Fig-Scan_1258keV}
\end{figure*}

\begin{figure*}
   \centering
    \includegraphics[width=0.85\textwidth]{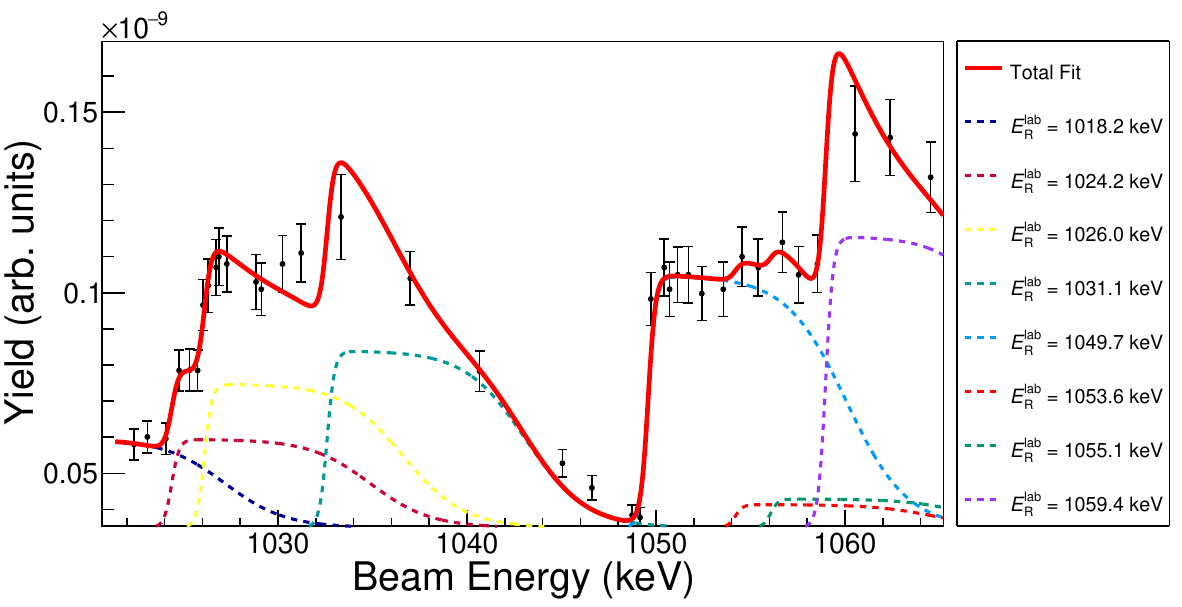}
  \caption{Resonance scan plot for the $E^{\rm{lab}}_{\rm{R}}$~=~1026.0, 1031.1, 1049.7, and 1059.4 keV resonances using the 889~keV secondary $\gamma$-ray transition. The red curve shows the total fit, while the dotted lines indicate contributions from individual resonances. 
  The target thickness obtained using Eq.~(\ref{equation-1}) is $\Delta E^{\rm lab} = 10.5(2)$~keV.
  }
  \label{fig:Fig-Scan_1026_1050keV}
\end{figure*}

First, the energy calibration of the accelerator was performed by measuring the front edge of the thick-target yield curve from the $E^{\rm{lab}}_{\rm{R}}$~=~992~keV resonance in the $^{27}$Al$(p,\gamma)^{28}$Si reaction~\cite{antilla1977use,keinonen1976standard}. 
Using the thick-target yield information of \citet{antilla1977use}, the detection efficiencies, $\eta$, were determined at various $\gamma$ energies (see Sec.~\ref{subsection:Detection_efficiency}).

In the second step, the $^{45}$Sc target was used to measure the thick-target yields over the $E^{\rm{lab}}_{\rm{R}}$~=~917.4, 1026.0, 1031.1, 1049.7, 1059.4, 
and 1257.5~keV resonances. A resonance scan was performed by observing the 889~keV secondary $\gamma$-ray transition with the HPGe in close geometry (\textit{d} = 1.7(2)~cm), where \textit{d} is the distance between the target holder and the end cap of the detector.
Measurement runs, typically lasting 1--3~h, were then carried out on the plateau of each resonance scan in far geometry (\textit{d} = 25.4(2)~cm), ensuring $<$5\% statistical uncertainty in the counts of the 889~keV $\gamma$-ray peak, to evaluate the yield and hence the strength of these resonances.

In the last step, the scan was done starting from the 917.4~keV resonance, going down to as low as 526~keV in steps of $\approx$10~keV.
The 889~keV secondary $\gamma$-ray transition was monitored using the HPGe detector placed in close geometry (\textit{d} = 1.7(2)~cm).

\section{Analysis}
\label{sec:Analysis}

\subsection{$\gamma$-ray spectra}
All resonances studied in this experiment de-excite via successive $\gamma$-ray transitions through the first excited state ($E_x = 889.286(3)$~keV) of $^{46}$Ti, ending with the emission of a secondary $\gamma$ ray at $E_\gamma = 889.277(3)$~keV. 
For the present analysis, only the 889~keV $\gamma$ ray is used to determine the yield $Y$. To determine the net counts in the 889~keV peak, the gross area was obtained by integrating the counts within the region of interest where the transition is visible, while the background was estimated using two adjacent sidebands on either side of the peak, following the integral method described in~\cite{gilmore}. A $\gamma$-ray spectrum for the $E^{\rm{lab}}_{\rm{R}} = 1049.7$~keV resonance is shown in Fig.~\ref{fig:Gamma_ray_1050keV}.

\subsection{Resonance scan}
\label{subsection:Resonance scan}
Resonance scans show the $\gamma$ yield from the resonance integrated over the target thickness. The shape of each scan reflects both the distribution of target material and the Breit-Wigner profile of the narrow resonance
~\cite{1948RvMP...20..236F,rolfs1990r}. The resonance strength is determined from the yield on the top of the scan.

All resonance scans were carried out by varying the proton beam energy and recording the yield of the  
889~keV $\gamma$ transition.
Each of these scans was fit with an empirical function~\cite{scottthesis2014,luna2015resonance}\footnote{This empirical formula, which is analytically simpler to implement, is derived from the Breit-Wigner formalism~\cite{1948RvMP...20..236F,rolfs1990r}.}
\begin{equation}
    f(E) = h \frac{1}{[1 + \mathrm{exp}(\frac{E^{\rm{lab}}_{\mathrm{R}} - E}{\delta_L})]}   \frac{1}{[1 + \mathrm{exp}(\frac{E - E^{\rm{lab}}_{\mathrm{R}} - \Delta E^{\rm{lab}}}{\delta_R})]},
\label{equation-1}
\end{equation}
where $h$ is the height of the plateau, and $E^{\rm{lab}}_{\mathrm{R}}$, $E$, and $\Delta E^{\rm{lab}}$ are the resonance energy, beam energy, and target thickness in the laboratory frame, respectively.
$\delta_{L}$ and $\delta_{R}$ are empirical parameters that characterize the steepness of the rising and falling edges of the plateau, respectively. The parameter $\delta_{L}$ includes contributions from the beam energy resolution, Doppler broadening, and the intrinsic resonance width, while $\delta_{R}$ incorporates these same initial effects but is predominantly governed by energy straggling of the beam and target non-uniformity.
The resonance scans for all resonances measured in this work are discussed individually in the following subsections, with additional details of the fitting procedure provided in Section I in~\cite{Sidhu_supplement}.

It should be noted that the scans of the resonances at 917.4, 1026.0, 1031.1, 1049.7, 1059.4 and 1257.5~keV reported in this work were also performed by \citet{Dubois1963} in 1963, but using a thinner target ($\approx$13 $\mu$g/cm$^2$). Resonances were measured in the range 873--1401.5~keV, and the relative intensities, corresponding to the thick-target yield of $\gamma$-rays with energies greater than 1.5~MeV, are shown in Fig.~2 and given in column 4 of Table~I in \citet{Dubois1963}.
In that work, to establish a consistent intensity scale for the resonances, the intensity of the thick-target yield from the 992~keV resonance of the $^{27}$Al($p,\gamma$)$^{28}$Si reaction, measured under the same conditions, was set to 10,000 units. When discussing our resonance scan, comparisons and references will therefore be made to Fig.~2 of~\cite{Dubois1963} and Table~I of~\cite{Dubois1963}.

\begin{figure*}
   \centering
    \includegraphics[width=0.85\textwidth]{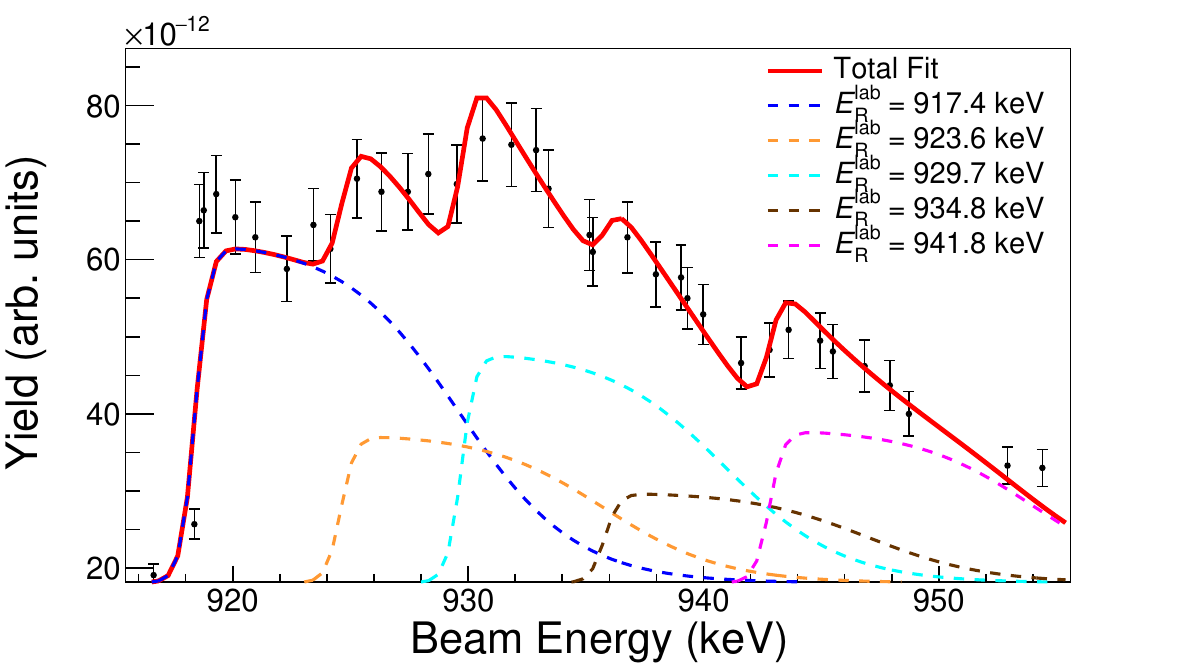}
  \caption{Resonance scan plot for the $E^{\rm{lab}}_{\rm{R}}$~=~917.4~keV resonance using the 889~keV secondary $\gamma$-ray transition. The red curve shows the total fit, while the dotted lines indicate contributions from individual resonances.  
  The target thickness obtained using Eq.~(\ref{equation-1}) is $\Delta E^{\rm lab} = 11.2(2)$~keV. 
  }
  \label{fig:Fig-Scan_917keV}
\end{figure*}

\subsubsection{$E^{\rm{lab}}_{\rm{R}}$ = 1257.5~keV resonance}
The scan of the 1257.5~keV resonance is shown in Fig.~\ref{fig:Fig-Scan_1258keV}. This resonance, identified as number 95 in Fig.~2 and Table~I of \citet{Dubois1963}, has the highest intensity of the reported resonances. 
Several nearby weaker resonances at 1248.6, 1252.4, 
1259.8, 1262.7, 1264.4, 1268.1, and 1272.2~keV (numbers 93--94 and 96--100, respectively, in Fig.~2 and Table~I of \cite{Dubois1963}) also contribute.
The individual contributions from each resonance are shown by the dotted lines in Fig.~\ref{fig:Fig-Scan_1258keV}.
The steepest increase in yield with increasing beam energy, as shown in Fig.~\ref{fig:Fig-Scan_1258keV}, corresponds to the 1257.5~keV resonance.
The strength was measured using a long run (taken at far geometry) acquired on the plateau region following the steep rise.

\subsubsection{$E^{\rm{lab}}_{\rm{R}}$ = 1026.0, 1031.1, 1049.7 and 1059.4~keV resonances}
\label{subsubsection:1026-scan}
In the beam energy range 1020--1063~keV, four strong resonances have been reported at 1026.0, 1031.1, 1049.7 and 1059.4~keV, corresponding to numbers 40, 41, 45 and 48 in Fig.~2 and Table~I of \citet{Dubois1963}.
Several nearby weaker resonances at 1018.2, 1022.8, 1024.2, 1037.2, 1041.1, 1044.3, 1053.6 and 1055.1~keV (numbers 37--39, 42--44, and 46--47 in Fig.~2 and Table~I of \citet{Dubois1963}) are also present. Among these, the contributions from the 1022.8, 1037.2, 1041.1 and 1044.3~keV resonances are negligible due to their low relative intensities and are not visible in the scan. The contribution from each resonance is indicated by the dotted lines in Fig.~\ref{fig:Fig-Scan_1026_1050keV}. The scan shows a clear increase in yield with increasing beam energy corresponding to the 1026.0, 1031.1, 1049.7 and 1059.4~keV resonances. 

During the experiment, the primary aim of this scan was to investigate the energy region and determine the strengths of the 1026.0 and 1049.7~keV resonances, for which long runs were taken on the plateau in far geometry. During the analysis, however, the 1031.1 and 1059.4~keV resonances were found to be clearly visible in the scan. Their strengths were therefore extracted from runs taken on the plateau in the close-geometry configuration.

\subsubsection{$E^{\rm{lab}}_{\rm{R}}$ = 917.4~keV resonance}
In the low-energy region, the 917.4~keV resonance is among the strongest resonances. 
The scan of this resonance (identified as number 11 in Fig.~2 and Table~I in \citet{Dubois1963}), shown in Fig.~\ref{fig:Fig-Scan_917keV}, was performed in the beam energy range 915--960~keV. 
Several nearby weaker resonances at 923.6, 929.7, 933.0, 934.8, 937.1, 941.8, 943.4, 946.4, 949.1, 953.8, 956.5 and 958.5~keV (numbers 12--23 in Fig.~2 and Table~I of \citet{Dubois1963}) are also present. Among these, the contributions from the 923.6, 929.7, 934.8 and 941.8~keV resonances are visible due to their relatively high intensities, while the others are not visible in the scan due to their negligible contributions.
The contribution from each resonance is indicated by the dotted lines in Fig.~\ref{fig:Fig-Scan_917keV}.
The steepest increase in yield with increasing beam energy, as shown in Fig.~\ref{fig:Fig-Scan_917keV}, corresponds to the 917.4~keV resonance.
The strength was measured from a long run (taken at far geometry) acquired on the plateau region following the steep rise.
To monitor possible target degradation, the scan was repeated three times, and the scans revealed no measurable degradation (see Fig. 1 in~\cite{Sidhu_supplement}).
\begin{figure*}
   \centering
    \includegraphics[width=0.85\textwidth]{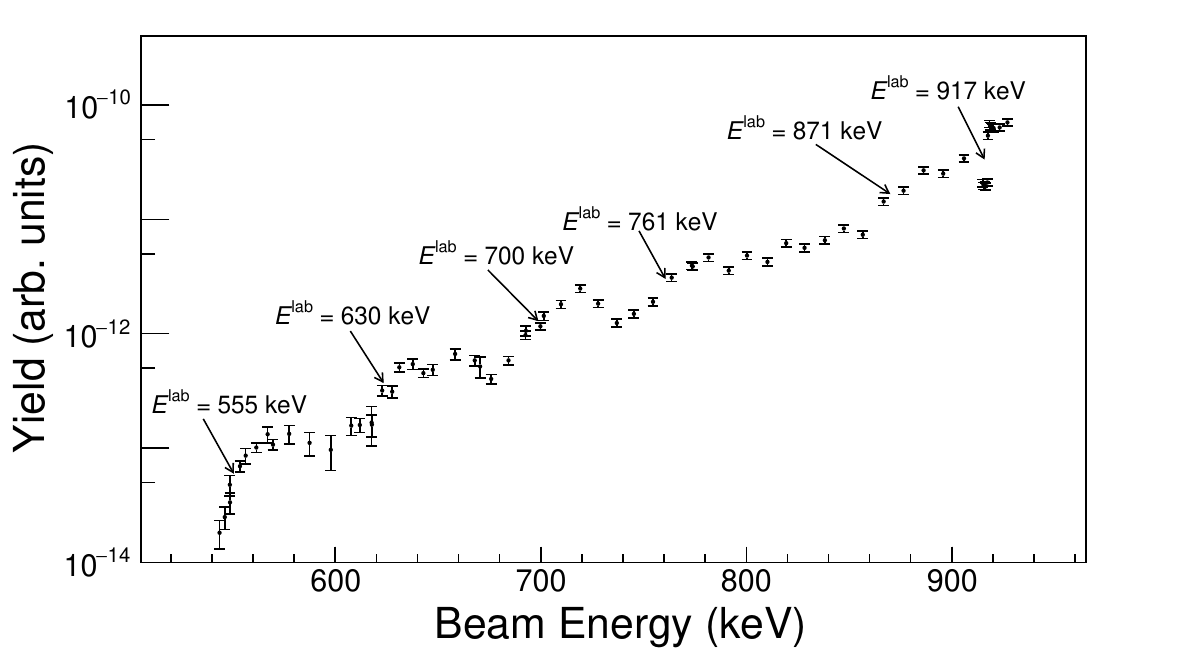}
  \caption{Resonance scan plot starting from the $E^{\rm{lab}}_{\rm{R}}$~=~917.4~keV resonance and going down to a proton beam energy of 526~keV using the 889~keV secondary $\gamma$-ray transition. The arrows indicate approximately the midpoint of the steep rise with increasing beam energy observed in the yield scan. }
  \label{fig:Fig-Scan_917keV_low}
\end{figure*}

\subsubsection{Resonances below $E^{\rm{lab}}_{\rm{R}}$ = 917.4 keV}
\label{subsubsection:below917keV}
In the low-energy region below 917.4~keV, a resonance scan was performed, as shown in Fig.~\ref{fig:Fig-Scan_917keV_low}. The scan started at the 917.4 resonance and extended down to 526~keV, with the beam energy reduced in steps of $\approx$10 keV.
The scan shows broad increases in yield extending over several tens of keV. 

At the beam energies in the present work, the proton ($S_p = 10344.8(7)$ keV), $\alpha$ ($S_\alpha = 8004.7(4)$ keV), and $\gamma$ emission channels are open, while the neutron emission channel remains closed ($S_n = 13189.2(9)$ keV).
To check if these structures 
arise from intrinsically broad resonances, upper limits on the partial widths $\Gamma_p$, $\Gamma_\alpha$, and $\Gamma_\gamma$ were calculated. 

For the calculation of the proton and $\alpha$ widths, the single-particle formalism~\cite{lane1960reduced,iliadis1997proton} was used, in which the partial width for a given level $\lambda$ and channel $c$ is given as
\begin{equation}
    \Gamma_{\lambda c} = 2 P_c \gamma^2_{\lambda c} = 2 P_c \frac{\hbar^2}{\mu R^2} \theta^2_{\lambda c},
\label{equation-2}
\end{equation}
where $\gamma^2_{\lambda c}$ and $\theta^2_{\lambda c}$ are the reduced width and dimensionless
reduced width, respectively, $\mu$ is the reduced mass, $R = r_0(A_p^{1/3}+A_t^{1/3})$ is the channel radius with $r_0 = 1.25$ fm, where $A_p$ and $A_t$ are the mass numbers of the projectile and target nuclei, respectively, and $P_c (E)$ is the Coulomb barrier penetrability. 
For a single-particle state, the dimensionless reduced width is $\theta^2_{\lambda c} = C^2S \theta^2_{\rm sp}$,  
where $C$ is the isospin Clebsch-Gordan coefficient, $S$ is the nucleon spectroscopic factor, and $\theta^2_{\rm sp}$ is the dimensionless single-particle reduced width. For the upper-limit estimates, a value of $C^2S \theta^2_{\rm sp} = 1$, corresponding to the Wigner single-particle limit, was used.
The penetrabilities $P_c(E)$ were calculated using \cite{penetrability}.
Assuming $s$-wave proton ($l = 0$) emission, corresponds to $f$-wave ($l = 3$) $\alpha$ emission in order to conserve angular momentum and parity\footnote{For $l_p = 0$ proton emission, the compound nucleus states in $^{46}$Ti have $J^\pi = 3^-$ or $4^-$, considering $J^\pi(^{45}\mathrm{Sc}) = 7/2^-$ and $J^\pi_p = 1/2^+$. For $\alpha$ emission to $^{42}$Ca($0^+$) + $\alpha(0^+)$ while conserving $J^\pi$, the allowed orbital angular momenta are $l_\alpha = 3$ or 5. Since low-spin states dominate in the astrophysical energy region, 
$l_\alpha = 3$ was used.}. 
Considering broad resonances at the lowest and highest energy structures, $E^{\rm c.m.} = 543$ keV ($E^{\rm lab} = 555$ keV) and $E^{\rm c.m.} = 852$ keV ($E^{\rm lab} = 871$ keV), the partial widths are $\Gamma_p \approx 0.16$ eV and $\Gamma_\alpha \approx 0.10$ eV at 543 keV, and $\Gamma_p \approx 38.24$ eV and $\Gamma_\alpha \approx 0.78$ eV at 852 keV, respectively\footnote{Assuming $s$-wave ($l = 0$) $\alpha$ emission,, $\Gamma_\alpha \approx 0.98$ eV and 7.11 eV and at 543 keV and 852 keV, respectively.}.

For the radiative width ($\Gamma_\gamma$), assuming a state at $E_x = 11.245$ MeV ($E^{\rm c.m.} = 900$ keV) in $^{46}$Ti decays predominantly via an $E1$ transition to the first excited state at 889~keV ($E_\gamma = 10.36$ MeV), a Weisskopf single-particle estimate~\cite{weisskopf1951radiative,blatt2012theoretical} gives $\Gamma_\gamma \approx 0.97$~keV, using $\Gamma_W(E1) = 0.068 \times A^{2/3} E_\gamma^3$ with $E_\gamma$ (in MeV).
However, in $fp$-shell nuclei, electric dipole transitions are significantly hindered relative to Weisskopf single-particle strengths~\cite{endt1979strengths}. More realistic $\Gamma_\gamma$ is therefore typically obtained by considering alternative multipolarities. For an $M1$ transition, the Weisskopf estimate yields $\Gamma_W(M1) = 0.021 \times E_\gamma^3 \approx 23.32$~eV, while for an $E2$ transition $\Gamma_W(E2) = 4.9 \times 10^{-8} A^{4/3} E_\gamma^5 \approx 0.96$~eV. Experimental systematics for $^{46}$Ti indicate that $M1$ transitions are also relatively weak, whereas $E2$ transitions exhibit enhanced strengths by factors of $\approx$ 12--52 compared to Weisskopf estimates~\cite{endt1979strengths}.

Even under the most optimistic assumptions, the total width, $\Gamma = \Gamma_\gamma + \Gamma_p + \Gamma_\alpha$, therefore remains below 0.1~keV; this check was performed for all observed structures. 
This demonstrates that the observed structure features, whose yield increases extend over several tens of keV, cannot be explained by intrinsically broad resonances with widths of less than 0.1~keV.
Given high-level density in this energy region (see Fig.~\ref{fig:Fig-Level_Scheme}) and the limited experimental beam energy resolution ($\approx$10 keV) of the scan, these structures are interpreted as the combined effect of multiple unresolved narrow resonances.
Under these conditions, any attempt to fit the individual features in Fig.~\ref{fig:Fig-Scan_917keV_low} would not be physically meaningful; therefore, only the experimental data points are shown without fitted curves.

For each observed structure, an effective energy is assigned, defined as the midpoint of the steep rise in yield. The corresponding strength is reported as an integrated resonance strength for that structure, determined from runs taken at the plateau in the close-geometry configuration.
It should be noted that this effective energy may not represent a single resonance energy, as each structure may comprise several unresolved resonances. Nevertheless, for the purpose of reaction rate calculations, each structure is treated as an effective single resonance, with the assigned effective energy taken as the resonance energy in the laboratory frame and the integrated strength taken as the corresponding resonance strength.
Further details on the extraction of resonance energies and strengths are provided in the following sections.

\subsection{Detection efficiency}
\label{subsection:Detection_efficiency}

\begin{figure}
   \centering
    \includegraphics[width=0.5\textwidth]{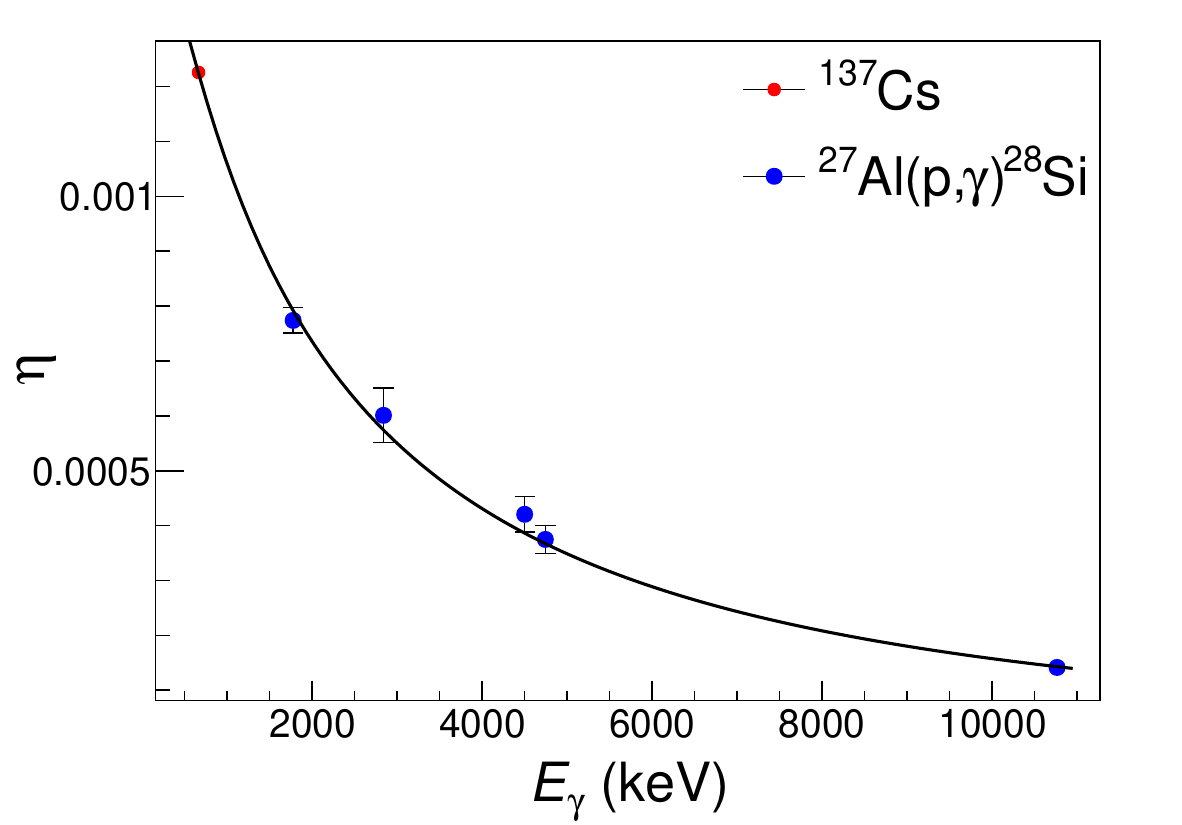}
  \caption{Absolute efficiency curve for the HPGe detector in far geometry. The
measured $\gamma$-ray peak efficiencies were fitted with an exponential function (Eq.~(\ref{equation-2})) indicated by the solid line.}
  \label{fig:Fig-HPGe}
\end{figure}

To determine the absolute detection efficiencies, $\eta$, for various $\gamma$-ray energies, a long run was taken on the plateau of the $E^{\rm{lab}}_{\rm{R}}$~=~992~keV resonance in the $^{27}$Al$(p,\gamma)^{28}$Si reaction, with the HPGe detector positioned in far geometry (\textit{d} = 25.4(2)~cm). 
The far geometry was chosen to minimize the summing effects.
To evaluate $\eta$, counts for the specific $\gamma$-ray were determined from the experimental spectrum and then normalized to the expected counts, 
based on the average yield of (1.08 $\pm$ 0.06)$\times$10$^{-9}$ 1778.9~keV $\gamma$-rays per incident proton, as reported by~\citet{keinonen1976standard}, and~\citet{paine1979p}, and the branching ratios provided in Table~I of~\citet{antilla1977use}.
As shown in Fig.~\ref{fig:Fig-HPGe}, the $\eta$ values obtained for different $E_\gamma$ were then fitted using the function~\cite{imbriani2005s}
\begin{equation}
    \eta(E_\gamma)  = \mathrm{e}^{a + b \mathrm{log}(E_\gamma) + c\mathrm{log^2}(E_\gamma) + d\mathrm{log^3}(E_\gamma) },
\label{equation-3}
\end{equation}
where \textit{a}, \textit{b}, \textit{c}, and \textit{d} are fitting parameters.
Using the above function and the obtained parameters from the fit, $\eta (E_\gamma)$ was calculated for a given $\gamma$-ray energy at $d$ = 25.4 cm. 

However, for the strength measurements of the resonances, runs were taken with the detector positioned at close geometry (\textit{d} = 1.7 cm). To obtain $\eta (E_\gamma)$ (\textit{d} = 1.7 cm), $\eta (E_\gamma)$ (\textit{d} = 25.4 cm) was scaled by a factor of $\eta_{\rm{close}}$/$\eta_{\rm{far}}$ = 30.2 $\pm$ 0.1.
To calculate $\eta_{\rm{close}}$/$\eta_{\rm{far}}$, multiple long runs were performed using a $^{137}$Cs source with the HPGe detector in close (1.7~cm) and far (25.4~cm) geometry. The $^{137}$Cs source emits a single primary $\gamma$ ray at 661 keV, and thus, there is no summing effect at any distance \textit{d}.

\begin{table}
\begin{ruledtabular}
\caption{Systematic uncertainty for the determination of the resonance strength $\omega \gamma$.} 
\begin{tabular}{ l r }                    
Uncertainty source & Contribution  \\   \hline            
Beam current   & 3\% \\
Stopping power   & 1.7\% \\
$\gamma$-ray detection efficiency   & 6.5\% \\
$\gamma$-ray summing (for close geometry runs)  & 15\% \\
Branching Ratio of the 889 keV gamma  & 10\% 
\end{tabular}
\label{tab:table1}
\end{ruledtabular}
\end{table}

\subsection{Summing effect}

All scans were performed in the close-geometry configuration. Long runs on the plateau for the 917.4, 1026.0, 1049.7 and 1257.5~keV resonances were, however, taken in far geometry, while long runs for the remaining resonances were taken in close geometry and therefore require correction for summing-out effects.

To determine the summing correction factor, runs were taken in both close and far geometry for the 917.4, 1026.0, 1049.7 and 1257.5~keV resonances. The yield of the 889~keV $\gamma$-ray was calculated and corrected for detection efficiency. The average ratio of the 889~keV yields (far to close geometry) was 1.68, and this factor was applied to runs taken in close geometry for the determination of resonance strengths.
The associated uncertainty was estimated from the maximum variation in this ratio (15\%). This correction and its uncertainty were applied only to the runs taken in close geometry.

\subsection{Branching ratio}
\label{sec:branching_ratio}

For the entire study of the $^{45}$Sc($p,\gamma$)$^{46}$Ti reaction in this work, the 889~keV secondary $\gamma$ ray was used. The branching ratio of the 889 keV transition from the first excited state to the ground state is reported to be 100\% with an uncertainty of 10\% \cite{Dubois1963}, which has been included in the overall error budget\footnote{Using \texttt{TALYS~v2.0}~\cite{Koning2023}, the calculated partial $^{45}$Sc($p,\gamma$)$^{46}$Ti cross section populating the 889~keV $2^+ \rightarrow 0^+$ transition accounts for more than 96\% of the total cross section over the entire energy range investigated in the present work, consistent with the branching ratio adopted from~\cite{Dubois1963}.}.

\subsection{Angular distributions}
The angular distributions can be expressed as a series of Legendre polynomials\begin{equation}
   W(\theta) = \sum_n a_n Q_n P_n(\mathrm{cos}\theta) \approx 1 + a_2 Q_2 P_2(\mathrm{cos}\theta),
\label{eq:equation-ad}
\end{equation}
where $a_n$ are the Legendre coefficients, determined by the spin and angular-momentum coupling and the $\gamma$-ray transition multipolarity, and $Q_n$ are attenuation factors that depend on the detector-target geometry~\cite{rose1953analysis}. As shown in Fig.~\ref{fig:Fig-Setup}, the HPGe detector is positioned at $45^\circ$ with respect to the beam direction.
Due to the large size of the HPGe detector, the effective detection angle can be safely assumed to be $\theta = 45^\circ \approx 55^\circ$ for both the close and far geometries. So angular distribution effects are negligible, and thus, no $W(\theta)$ correction is required, irrespective of the magnitude of the $a_2$ coefficient. 

\subsection{Resonance energy}
\label{sec:resonance_energy}

For the higher-energy resonances at $E^{\rm lab}_{\rm R}$ = 917.4, 1026.0, 1031.1, 1049.7, 1059.4 and 1257.5~keV, the resonance energies are taken from~\cite{Dubois1963} and are consistent with the present measurements. An uncertainty of 1~keV, as reported in~\cite{Dubois1963}, is included in this work. The resonance energy values are provided in Table~\ref{tab:table2}. 

In the low-energy region, where broad yield structures are observed, an effective energy is assigned to each structure and is defined as the midpoint of the leading edge of the yield curve, which corresponds to the resonance energy in the laboratory frame under thick-target conditions~\cite{1948RvMP...20..236F}.
It should be noted that this does not correspond to a single resonance energy, as each structure may be interpreted as a combination of multiple unresolved resonances.
To estimate the associated uncertainty, the structure exhibiting the largest energy spread is considered. The uncertainty is taken as the difference between the highest and lowest beam energies over which the yield varies across the structure. This results in an uncertainty of $\pm 25$ keV.
The resonance energy values are provided in Table~\ref{tab:table3}.

\subsection{Resonance strength}
\label{sec:resonance_strength}

For the thick-target yield condition~\cite{1948RvMP...20..236F}, the resonance strength, $\omega \gamma$, is determined from the experimental yield, \textit{Y}, 
and is given by
\begin{equation}
    \omega \gamma = \frac{2}{\lambda_{\rm r}^2} Y \epsilon_{\rm r} \frac{M}{m+M},
\label{eq:equation-4}
\end{equation}
where $\lambda_{\rm r}$, is the de Broglie wavelength at the resonance energy in the center-of-mass frame, $\epsilon_{\rm r}$ is the effective stopping power at the resonance energy in the laboratory frame (calculated using \texttt{SRIM-2013}~\cite{ziegler2010srim}), and $m$ and $M$ are the projectile and target masses, respectively.
For all the resonances reported in this work, resonance width, $\Gamma$, is significantly smaller than the energy loss (in the center-of-mass frame) ($\Gamma$ $\textless \textless$ $\Delta E_{\rm c.m.}$) and hence the thick-target condition is well satisfied.
Yield is calculated using
\begin{equation}
    Y = \frac{N e}{\eta \; W(\theta) \; \mathrm{BR} \; q } ,
\label{eq:equation-5}
\end{equation}
where $N$ is the counts from the 889~keV secondary $\gamma$ rays, $e$ is the elementary charge ($e$ = $1.602 \times 10^{-19}$ C), $q$ is the total
accumulated charge from the incident proton beam, $\eta$, $W(\theta)$, and BR are 
the detection efficiency, angular distribution, and branching ratio 
of the emitted $\gamma$, respectively.

As evident from the scan plots, multiple resonances are present within a single scan. The $Y$ for a given resonance is calculated using
\begin{equation}
    Y = Y_{\mathrm{plateau}} - Y_{\mathrm{baseline\,below\,resonance}},
    \label{eq:equation-6}
\end{equation}
where $Y_{\mathrm{plateau}}$ represents the yield on the plateau of the resonance of interest, and $Y_{\mathrm{baseline \, below \, resonance}}$ corresponds to the yield at the baseline preceding the steep rise.
An example of a baseline-below-resonance run together with a plateau run is shown for the 1257.5 keV resonance in Fig.~2 of~\cite{Sidhu_supplement}.

For resonances above and below 900~keV, the above general analysis procedure is applied.
For resonances above 900~keV, individual resonance strengths are reported, whereas for those below 900~keV, integrated resonance strengths are given due to the presence of unresolved structures. It is to be noted that for resonances above 900~keV, the thin-target data of~\cite{Dubois1963} provide useful guidance on the number of resonance contributions required to fit the present data. In contrast, for resonances below 900~keV, no such prior information is available.
The extracted resonance strengths for $E^{\rm lab}_{\rm R} >$ 900~keV are listed in Table~\ref{tab:table2}, while for $E^{\rm lab}_{\rm R}  < $ 900~keV the corresponding integrated strengths are provided in Table~\ref{tab:table3}, together with their statistical and systematic uncertainties.

The total uncertainty is dominated by systematic contributions, the individual components of which are summarized in Table~\ref{tab:table1}. It should be noted that all baseline runs 
were taken in the close-geometry configuration. Consequently, the overall systematic uncertainty on all strength values is obtained by adding the contributions in quadrature, resulting in a total uncertainty of $\approx$20\%.

\begin{table*}
\begin{ruledtabular}
\caption{Resonance parameters for the resonances above 900~keV studied in the present work.} 
\begin{tabular}{ c c c }                    
$E^{\rm{lab}}_{\rm{R}}$ (keV) & $E_{x}$ (keV)\footnote{For the calculation of $E_{x}$, $S_p$ = 10344.8(7) keV~\cite{WU20001} is used.} & $\omega \gamma$ (eV) (present work)  \\   \hline            
1257.5(10)~\cite{Dubois1963} & 11574.7(12)  & 0.477 $\pm$ 0.005$_{\rm stat}$ $\pm$ 0.095$_{\rm syst}$ \\
1059.4(10)~\cite{Dubois1963} & 11381.0(12)  & 0.176 $\pm$ 0.015$_{\rm stat}$ $\pm$ 0.035$_{\rm syst}$ \\
1049.7(10)~\cite{Dubois1963} & 11371.5(12)  & 0.362 $\pm$ 0.003$_{\rm stat}$ $\pm$ 0.072$_{\rm syst}$ \\
1031.1(10)~\cite{Dubois1963} &  11353.3(12)  & 0.058 $\pm$ 0.011$_{\rm stat}$ $\pm$ 0.012$_{\rm syst}$ \\
1026.0(10)~\cite{Dubois1963} & 11348.3(12)  & 0.124 $\pm$ 0.006$_{\rm stat}$ $\pm$ 0.025$_{\rm syst}$ \\
917.4(10)~\cite{Dubois1963} & 11242.1(12)  & 0.137 $\pm$ 0.003$_{\rm stat}$ $\pm$ 0.027$_{\rm syst}$ \\                    
\end{tabular}
\label{tab:table2}
\end{ruledtabular}
\end{table*}

\begin{table}
\begin{ruledtabular}
\caption{Resonance parameters for the resonances below 900~keV studied in the present work.} 
\begin{tabular}{ c c}                    
$E^{\rm{lab}}$ (keV) &   $\sum \omega \gamma$ (eV)   \\   \hline            
871(25)  &   0.0982 $\pm$ 0.0003$_{\rm stat}$ $\pm$ 0.0196$_{\rm syst}$ \\
761(25) &    0.0153 $\pm$ 0.0006$_{\rm stat}$ $\pm$ 0.0031$_{\rm syst}$ \\
700(25) &    0.0086 $\pm$ 0.0003$_{\rm stat}$ $\pm$ 0.0017$_{\rm syst}$ \\
630(25) &  0.0014 $\pm$ 0.0002$_{\rm stat}$ $\pm$ 0.0003$_{\rm syst}$ \\
555(25)  &   0.00042 $\pm$ 0.00008$_{\rm stat}$ $\pm$ 0.00008$_{\rm syst}$ \\
                       
\end{tabular}
\label{tab:table3}
\end{ruledtabular}
\end{table}

\section{Transfer reaction data}
\label{sec:Transfer_reaction_data}

The lowest proton beam energy investigated in the present work corresponds to 526~keV. However, as discussed in Sec.~\ref{sec:Literature_Review}, two earlier proton transfer studies reported proton-unbound states at even lower excitation energies in $^{46}$Ti~\cite{broman1968level,broman1971the45sc}. Below 526~keV, seven proton-unbound levels have been reported, and in order to provide a more complete thermonuclear reaction rate, their resonance strengths were estimated using nuclear-structure predictions for medium-mass nuclei~\cite{BrownRae2014}.

The excitation energies of the seven lowest proton-unbound states were taken from~\citet{broman1968level} and are listed in Table~\ref{tab:table4}. In the subsequent study by~\citet{broman1971the45sc}, an orbital angular momentum transfer of $l_p = 3$ was reported for two of these states. 
For the remaining levels, the proton partial widths $\Gamma_p$ were estimated assuming $s$-wave proton capture ($l_p = 0$)\footnote{For higher values of $l_p$, the proton penetrability decreases, which in turn will reduce both $\Gamma_p$ and the corresponding $\omega \gamma$ values listed in Table~\ref{tab:table4}.}.

The proton partial widths were calculated using the single-particle formalism described in Eq.~\ref{equation-2}.
At excitation energies around 10~MeV in $^{46}$Ti, the single-particle strength is expected to be strongly fragmented over many compound-nuclear configurations, since this energy region lies well above the low-lying shell-model states and approaches the onset of the quasi-continuum. 
A nucleon spectroscopic factor of $S = 0.1$ was therefore used.
This value is consistent with shell-model calculations performed using the NuShellX@MSU code with the large-basis shell model (LBSM) interaction in the full $fp$ model space~\cite{BrownRae2014}. The uncertainties in the shell-model spectroscopic strengths are estimated to be in the 20\% range~\cite{Lee2006}; therefore, a 20\% uncertainty was assigned to the spectroscopic factor $S$ and, consequently, to the calculated $\Gamma_p$ values.
The isospin Clebsch-Gordan coefficient, $C^2$, accounts for the coupling between the target, projectile, and final-state isospins. 
For a $^{45}\mathrm{Sc}$ target with isospin $T_t = 3/2$ and a proton projectile with $T_p = 1/2$, the entrance channel can populate compound states in $^{46}\mathrm{Ti}$ with total isospin $T_f = 1$ or $T_f = 2$.
Since the $T_f = 2$ isobaric analog states lie at higher excitation energies ($E_x \approx 13~\mathrm{MeV}$), well above the current range of interest ($E_x = 10.374\text{--}10.782$~MeV), the observed resonances are assumed to be predominantly $T_f = 1$ states, which gives $C^2 = 0.75$ and therefore $C^2S = 0.075$.
This value is consistent with shell-model calculations performed using the NuShellX@MSU code with the LBSM interaction in the full $fp$ model space~\cite{BrownRae2014}.

The dimensionless single-particle reduced widths (as defined in Eq.~\ref{equation-2}) were taken as $\theta^2_{\rm sp} = 0.51$ for $l = 0$ and $\theta^2_{\rm sp} = 0.30$ for $l = 3$, corresponding to the average of the $\theta^2_{\rm sp}$ values for $A = 40$ and $A = 50$ listed in Table~III of~\cite{iliadis20262025}.
The penetrabilities $P_c$ were evaluated at $E^{\rm c.m.}_{\rm R}$ using~\cite{penetrability}\footnote{It should be noted that the $E_x$ (see column~1 in Table~\ref{tab:table4}) carry an uncertainty of $\pm 25$~keV, and that a variation of $\pm 25$~keV in the energy region $E^{\rm c.m.}_{\rm R} \ge 263$~keV can change the $P_c$ by up to a factor of 8--10. At the lowest three resonance energies, this sensitivity can be even more pronounced; however, the three lowest-energy resonances lie well outside the Gamow window for the scenarios considered here and therefore do not contribute to the reaction rate (see Sec.~\ref{sec:Reaction_Rate} and Fig.~\ref{fig:Fig-Rate_Ratio_Plot}).}.

The strength for a single resonance, $\omega \gamma$, is derived from the Breit-Wigner function description of an isolated resonance~\cite{BreitWigner1936} and is typically defined as
\begin{equation}
   \omega \gamma = \frac{2J + 1}{(2J_1 + 1)(2J_2 + 1)}\frac{\Gamma_p \Gamma_{\gamma}}{\Gamma},
\label{eq:equation-1.1}
\end{equation}
where $\Gamma_p$, $\Gamma_\gamma$, and $\Gamma$ are the proton partial, radiative and total, respectively~\cite{Fowler1967}, with $\Gamma = \Gamma_p + \Gamma_\gamma + \Gamma_\alpha$.
The statistical factor, $\omega$, is related to the spins $J_1$, $J_2$ and $J$ of the projectile, target, and intermediate state that corresponds to the resonance, respectively. 

As discussed in Sec.~\ref{subsubsection:below917keV}, electromagnetic transition strengths in $fp$-shell nuclei deviate from Weisskopf single-particle estimates. $E1$ transitions in $^{46}$Ti are strongly hindered (strengths typically reduced to $5\times10^{-5}$--$1.4\times10^{-3}$ of the Weisskopf single-particle estimates), while $E2$ transitions exhibit enhancement factors of approximately 12--52 relative to the Weisskopf values~\cite{endt1979strengths}. $M1$ and $M2$ transitions are also weaker than the corresponding Weisskopf single-particle estimates.
Assuming that the highest-lying state at $E_x = 10.782$ MeV in Table~\ref{tab:table4} decays to the first excited state at 889~keV via an $E2$ transition ($E_\gamma = 9.89$~MeV), the Weisskopf estimate gives $\Gamma_\gamma \approx 0.77$ eV, while using the lowest enhancement factor of 12 gives $\Gamma_\gamma \approx 9.18$~eV. This is substantially larger than the calculated $\Gamma_p$ listed in Table~\ref{tab:table4}.

At these energies, the $\alpha$ channel is energetically open and $\Gamma_\alpha >> \Gamma_p$\footnote{For example, the $s$-wave proton emission in the $E_{\text{R}}^{\text{lab}} = 394\text{ keV}$ resonance corresponds to an $f$-wave $\alpha$ emission. Assuming an upper-limit spectroscopic factor of $C^2S\theta^2_{\mathrm{sp}} = 1,$ 
$\Gamma_\alpha \approx 2.83 \times 10^{-2}\text{ eV}.$ While this upper limit may decrease depending on the $\alpha$ strength function, it remains larger than the $\Gamma_p$ value listed for the $394\text{ keV}$ resonance in Table~\ref{tab:table4}.}.   
However, both $\Gamma_p$ and $\Gamma_\alpha$ are much smaller than $\Gamma_\gamma$.
Therefore, in the low-energy regime considered here, $\Gamma_p$ $<< \Gamma_\alpha << \Gamma_\gamma$, and thus the total width $\Gamma$ $\approx$ $\Gamma_\gamma$, such that the resonance strength simplifies to  
\begin{equation}
   \omega \gamma \approx \frac{2J + 1}{(2J_1 + 1)(2J_2 + 1)} \Gamma_p. 
\label{eq:equation-1.2}
\end{equation}

Coupling the $^{45}$Sc ground state ($J_1^\pi = 7/2^-$) with the incident proton ($J_2^\pi = 1/2^+$) gives rise to two possible spin-parity assignments, $J^\pi = 3^-$ and $4^-$, for states populated through $l=0$ proton capture in $^{46}$Ti. For states populated through $l=3$ proton capture, the possible spin-parity values range from $0^+$ to $7^+$. A statistical factor of $\omega = 0.5$ was used with an associated uncertainty of 87.5\%, corresponding to the possible range of $\omega$ values from $1/16$ to $15/16$ for $J = 0$ and $J = 7$, respectively. This is consistent with the factor of 1.5--2 uncertainty obtained when low spectroscopic factors and poorly constrained spin assignments are sampled using lognormal distributions~\cite{longland2010charged}. The resulting resonance strengths using the transfer reaction measurement data are listed in Table~\ref{tab:table4}.

\begin{table*}
\begin{ruledtabular}
\caption{Resonance parameters for states below 526 keV calculated using the transfer reaction data information from~\cite{broman1968level,broman1971the45sc}.} 
\begin{tabular}{ c c c c  c}                    
$E_{x}$ (keV) &  $E^{\rm{lab}}_{\rm{R}}$ (keV) & $l_p$ & $\Gamma_p$ (eV)  & $ \omega \gamma$ (eV)   \\   \hline            
10374(25)~\cite{broman1968level}  & 30(26) & 0 & 34.75(6.95) $\times 10^{-44}$	 &	1.74(1.56) $\times 10^{-43}$ \\
10441(25)~\cite{broman1968level} & 98(26) & 0  & $13.45(2.69)\times 10^{-20}$	 &	$6.73(6.04)\times 10^{-20} $ \\
10526(25)~\cite{broman1968level} &  185(26) &3~\cite{broman1971the45sc} & $5.06(1.01)\times 10^{-15}$	 &	$2.53(2.27)\times 10^{-15} $ \\
10602(25)~\cite{broman1968level} &  263(26) &0& $21.16(4.23)\times 10^{-9}$	 &	$10.58(9.49)\times 10^{-9}$  \\
10670(25)~\cite{broman1968level}  & 332(26) & 0 & $18.52(3.71)\times 10^{-7}$	 &   $9.26(8.31)\times 10^{-7}$  \\
10730(25)~\cite{broman1968level} &  394(26) & 0 &$33.98(6.79)\times 10^{-6}	$&	$1.70(1.52)\times 10^{-5}$  \\
10782(25)~\cite{broman1968level}  & 447(26) & 3~\cite{broman1971the45sc} &$ 19.97(3.99)\times 10^{-8}$	 &	$9.99(8.96)\times 10^{-8}$  \\

\end{tabular}
\label{tab:table4}
\end{ruledtabular}
\end{table*}

\section{Thermonuclear Reaction Rate}
\label{sec:Reaction_Rate}

Assuming all 18 resonances, including the 11 directly measured resonances from the present work and the 7 resonances inferred from transfer-reaction data, are narrow and isolated, the total thermonuclear reaction rate for the $^{45}$Sc($p,\gamma$)$^{46}$Ti reaction was calculated using~\cite{Fowler1967, rolfs1990r} 
\begin{multline}
N_A \langle\sigma v\rangle \left[\frac{\rm{cm^3}}{\rm s\cdot mol}\right] = N_A  \langle\sigma v\rangle_{\rm{R}} = \frac{1.54 \times 10^{11}} {(\mu T_{9})^{3/2}}  \\ \times \sum_{i} (\omega \gamma \times 10^{-6})_i  \times \mathrm{exp}\left( \frac{-11.605 \times E^{\mathrm{c.m.}}_{{\rm{R}},i}}{1000 \times T_{9}}\right), 
\label{eq:rr_numberical}
\end{multline}
where $\mu$ is the reduced mass in atomic mass unit u, $\omega \gamma$ is the resonance strength in eV, and $E^{\mathrm{c.m.}}_{\rm{R}}$ is the resonance energy in the center-of-mass frame in keV.

\begin{figure}
   \centering
    \includegraphics[width=0.5\textwidth]{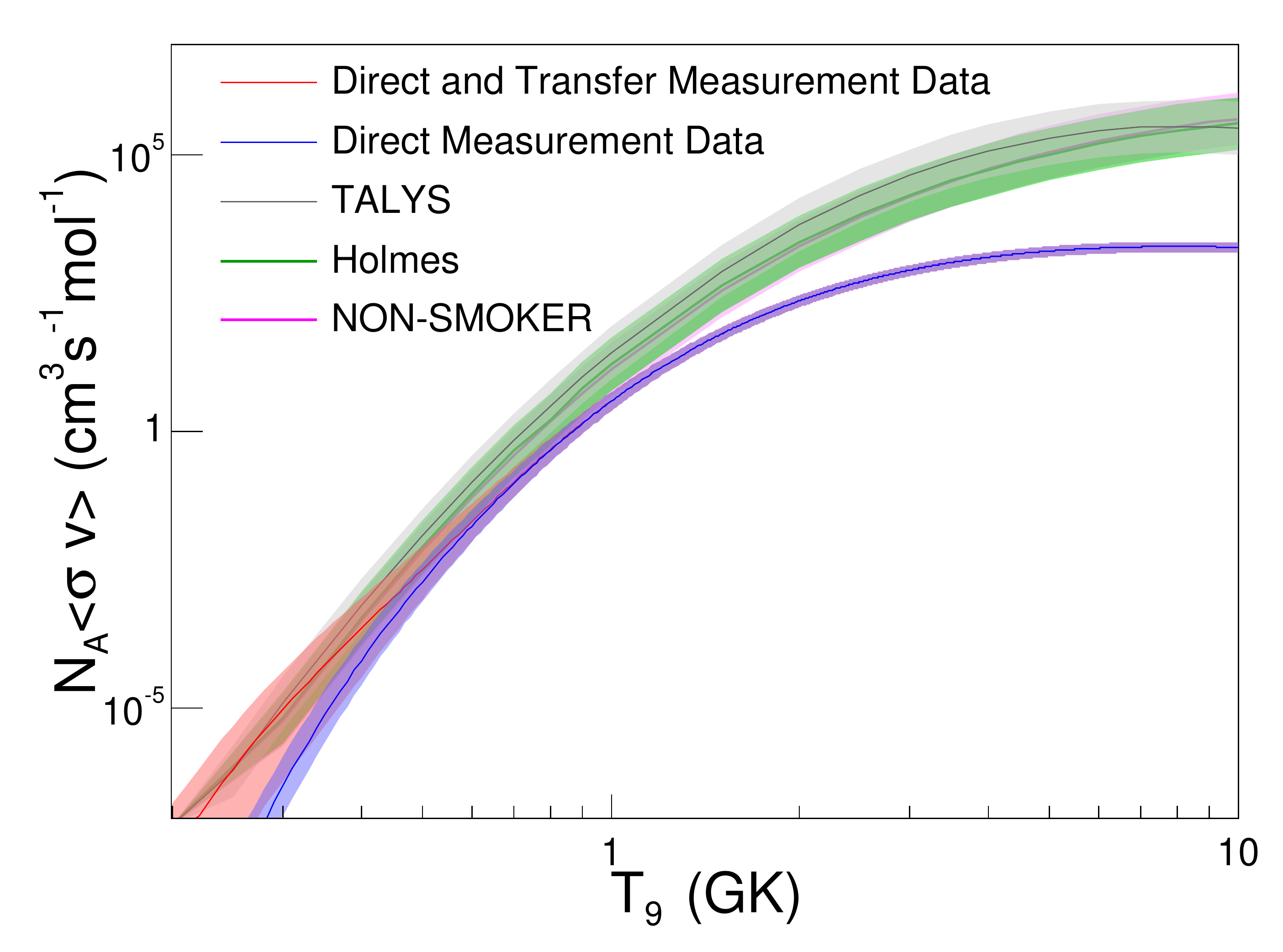}
  \caption{Reaction rate components for the $^{45}$Sc($p,\gamma$)$^{46}$Ti reaction as a function of temperature. The blue curve shows the total reaction rate obtained from the 11 resonances measured directly in the present work, whereas the red curve includes, in addition, the 7 low-energy resonances inferred from transfer reaction studies. The grey, green, and magenta curves correspond to calculations performed with \texttt{TALYS v2.0}, classical Hauser--Feshbach model of Holmes~\cite{Holmes1976}, and \texttt{NON-SMOKER}, respectively, with the magenta and green curves overlapping each other.}
  \label{fig:Fig-Rate_Plot}
\end{figure}

\begin{figure*}
   \centering
    \includegraphics[width=\textwidth]{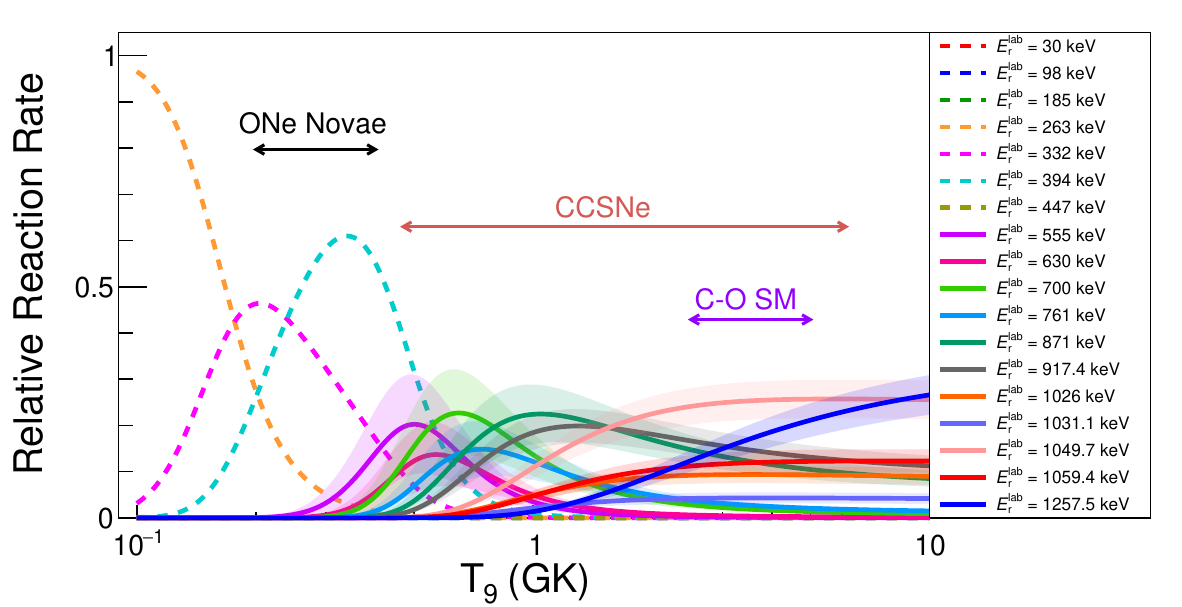}
  \caption{Contribution of individual resonances to the total thermonuclear reaction rate of the $^{45}$Sc($p,\gamma$)$^{46}$Ti reaction as a function of temperature. The dotted curves correspond to resonances inferred from transfer-reaction studies; since the reaction rate depends exponentially on $E^{\rm c.m.}_{\rm R}$ and these level energies carry an uncertainty of 25 keV that dominates the reaction rate, the resulting uncertainty becomes significant, and therefore no uncertainty band is shown. The remaining resonances were measured directly in the present work. }
  \label{fig:Fig-Rate_Ratio_Plot}
\end{figure*}

The total thermonuclear reaction rate for the $^{45}$Sc($p,\gamma$)$^{46}$Ti reaction is shown in Fig.~\ref{fig:Fig-Rate_Plot}, while the relative contributions of individual resonances are shown in Fig.~\ref{fig:Fig-Rate_Ratio_Plot}. The first (blue curve) includes only the 11 resonances measured directly in the present work. The second (red curve) additionally includes the contributions of 7 resonances inferred from transfer-reaction data listed in Table~\ref{tab:table4}\footnote{The $E^{\rm{lab}}_{\rm{R}}$ = 30, 98, and 185~keV resonances do not contribute over the temperature range 0.2--10~GK; thus, effectively only five of these seven resonances contribute to the reaction rate, as indicated by the dotted lines in Fig.~\ref{fig:Fig-Rate_Ratio_Plot}.}\footnote{The reaction rates corresponding to the blue and red curves are given in Tables~I and II, respectively, in~\cite{Sidhu_supplement}.}.
A difference between the two cases is observed at low temperatures ($T < 0.5$ GK), although it remains within the uncertainty bands. In this regime, the reaction rate is dominated by contributions from low-energy resonances inferred from transfer-reaction data (see dotted lines in Fig.~\ref{fig:Fig-Rate_Ratio_Plot}). At higher temperatures ($T > 0.5$ GK), the red and blue curves converge, indicating that the reaction rate is increasingly governed by the higher-energy resonances measured directly in the present work. This temperature range is relevant for core-collapse supernovae (CCSNe), oxygen and silicon burning, and C-O shell mergers (see the next section for further discussion).

For comparison, statistical model predictions are also shown. These calculations rely on standard assumptions for nuclear level densities, deformation effects, proton transmission coefficients~\cite{PhysRev.109.1243}, and the $\gamma$-ray strength function~\cite{ISAAK2019225}. The grey curve corresponds to calculations performed with \texttt{TALYS v2.0}~\cite{Koning2023} using default input parameters,
the green curve represents the classical Hauser-Feshbach model~\citet{Holmes1976}, which was used in an earlier evaluation of the reaction rate~\cite{1978ApJ...223..697S},
and the magenta curve shows results from the \texttt{NON-SMOKER} using default parameters~\cite{RAUSCHER2001,non-smoker}, with the green and magenta curves overlapping each other.
Standard Hauser-Feshbach calculations, such as those performed with \texttt{TALYS} and \texttt{NON-SMOKER}, typically reproduce measured cross sections within a factor of approximately 2--3 when the statistical model is applicable, and the relevant optical-model potentials, level densities, and $\gamma$-strength functions are reasonably constrained~\cite{Iliadis2001}. Accordingly, an uncertainty band corresponding to a factor of 3 above and below the statistical-model predictions is shown in Fig.~\ref{fig:Fig-Rate_Plot}.

For comparison with the statistical model calculations, the red curve should be considered.
This reaction rate matches the statistical-model predictions within uncertainties up to $T \approx 1.2$ GK. 
At higher temperatures ($T \gtrsim 1.2$ GK), the experimentally derived rate falls below the statistical-model predictions, since the present evaluation includes only experimentally identified resonances, whereas statistical models incorporate a much larger number of unresolved and unobserved states through averaged nuclear properties. Although the level density in $^{46}$Ti above the proton threshold is high (direct capture contributions are expected to be negligible), many additional states are either too weak to contribute individually or are effectively included within the statistical treatment of the compound nucleus.

\section{Discussion}
\label{sec:discussion}

The effective energy range over which nuclear reactions occur in stellar environments is expressed by the Gamow window~\cite{atkinson1929frage,gamow1938rate}. The center of the Gamow window, referred to as the Gamow energy, is given by
\begin{equation}
E_{0} = 0.122 \left( Z_{1}^{2} Z_{2}^{2} \mu T_{9}^{2} \right)^{1/3} \, \text{MeV},
\end{equation}
where $Z_1$ and $Z_2$ are the atomic numbers of the interacting nuclei, $\mu$ is the reduced mass in atomic mass units, and $T_9$ is the temperature in GK. The corresponding width of the Gamow window is given by
\begin{equation}
\Delta = 0.237 \left( Z_{1}^{2} Z_{2}^{2} \mu T_{9}^{5} \right)^{1/6} \, \text{MeV}.
\end{equation}

For oxygen-neon (ONe) novae temperatures ($T \approx 0.2$--0.4 GK~\cite{downen2013nuclear}), the corresponding Gamow window in the laboratory frame spans $E^{\rm lab} = 235$--667 keV. Within this energy range, the reaction rate is dominated by low-energy resonances inferred from transfer-reaction data (see dotted lines in Fig.~\ref{fig:Fig-Rate_Ratio_Plot}). These states are taken from the transfer-reaction studies~\cite{broman1968level,broman1971the45sc}. 
For these states, the uncertainty in the excitation energies is 25 keV, which significantly impacts the reaction rate.
Improved transfer-reaction measurements and $\gamma$-ray spectroscopy studies, providing more precise excitation energies as well as reliable angular-momentum transfer information and spin-parity assignments, are required to better constrain the reaction rate in this astrophysically important temperature regime for ONe novae.

For oxygen and silicon shell-burning temperatures ($T = 2.5$--5 GK~\cite{woosley1995evolution}),
relevant to C-O shell mergers, the Gamow window in the laboratory frame spans $E^{\rm lab} = 1044$–2657 keV. Within this range, the present study includes resonances up to 1257.5 keV. Further measurements of higher-lying, narrow, and isolated resonances are therefore required to more reliably constrain the reaction rate in this regime. Previous studies of this energy range~\cite{1978ApJ...223..697S} adopted a simplified statistical model approach~\cite{Holmes1976} for deriving the reaction rate at these higher temperatures. 

For CCSNe temperatures ($T = 0.47$--6.15 GK~\cite{hermansen2020reaction}), relevant to $^{47}$Sc production, the corresponding Gamow window extends from $E^{\rm lab} = 392$ to 4675 keV. All resonances measured in this work fall within this range. 
However, additional precise measurements are needed, particularly for resonances above 1257.5 keV and below 900 keV, in order to achieve a more complete determination of the reaction rate.

\section{Summary}
\label{sec:summary}
In summary, the first direct measurements of the resonance strengths for six low-energy resonances in the $^{45}$Sc($p,\gamma$)$^{46}$Ti reaction have been reported at $E^{\rm lab}_{\rm R} = 917.4$, 1026.0, 1031.1, 1049.7, 1059.4 and 1257.5 keV. In addition, integrated resonance strengths were determined for broad unresolved structures at $E^{\rm lab} = 555$, 630, 700, 761 and 871 keV. For all 11 resonances, the strengths were determined using the secondary 889 keV $\gamma$ transition.
For resonance energies below the lowest directly measured beam energy of 526 keV, available transfer-reaction data were used to estimate the resonance strengths based on the statistical assumptions on the various transition strengths. Combining the directly measured resonances with these low-energy constraints, an updated thermonuclear reaction rate for the $^{45}$Sc($p,\gamma$)$^{46}$Ti reaction was evaluated over the astrophysically relevant temperature range and compared with statistical model predictions. The results suggest that the level density in this mass range is sufficiently high, and the single-resonance-based reaction rate agrees well with the statistical model predictions within the given uncertainty range. 
This is in agreement with previous observations~\cite{Berg_2012} that, upon entering the $fp$ shell, the level density increases significantly and the discrete level structure becomes increasingly fragmented due to the larger number of available quantum configurations and excitation modes built upon them~\cite{brown2018endf}.

\begin{acknowledgments}
The authors would like to thank all members of the
Nuclear Science Laboratory at the University of Notre Dame
for their tremendous support. 
The authors express their thanks to A. Giambrone and S. Sullivan for performing \texttt{TALYS} calculations.
This research was supported by the National Science Foundation through Grant No.~PHY-2011890 and Grant No. PHY-2310059 (University of Notre Dame Nuclear Science Laboratory), Grant No.~PHY-1430152 (the Joint Institute for Nuclear Astrophysics - Center for the Evolution of the Elements), Grant No. OISE-1927130 (IReNA), the U.S. Department of Energy under Award Numbers DE-SC0026204 and DE-SC0023128 (CeNAM), the Royal Society International Exchanges Grant (IES\textbackslash R2\textbackslash232267), the Royal Society Research Grants (RGS\textbackslash R1\textbackslash 261005), and the Science and Technology Facilities
Council (STFC).
\end{acknowledgments}

\end{document}


\title{Supplemental Material for: First strength measurements of low-energy resonances in the $^{45}$Sc($p,\gamma$)$^{46}$Ti reaction and its astrophysical implications}

\author{R.~S.~Sidhu}
\email{ragan.sidhu@surrey.ac.uk}
\affiliation{School of Mathematics and Physics, University of Surrey, Guildford, GU2 7XH, United Kingdom}

\author{R.~J.~deBoer}
\affiliation{Department of Physics and Astronomy, University of Notre Dame, Notre Dame, Indiana 46556, USA}

\author{J.~Görres}
\affiliation{Department of Physics and Astronomy, University of Notre Dame, Notre Dame, Indiana 46556, USA}

\author{M.~Wiescher}
\affiliation{Department of Physics and Astronomy, University of Notre Dame, Notre Dame, Indiana 46556, USA}

\author{W.~N.~Catford}
\affiliation{School of Mathematics and Physics, University of Surrey, Guildford, GU2 7XH, United Kingdom}

\author{C.~Dembski}
\affiliation{Department of Physics and Astronomy, University of Notre Dame, Notre Dame, Indiana 46556, USA}

\author{C.~R.~Jones}
\affiliation{Department of Physics and Astronomy, University of Notre Dame, Notre Dame, Indiana 46556, USA}

\author{Ó.~E.~López-López}
\affiliation{Instituto de Física, Universidad Nacional Autónoma de México, 04510, Ciudad de México, Mexico}

\author{G.~Lotay}
\affiliation{School of Mathematics and Physics, University of Surrey, Guildford, GU2 7XH, United Kingdom}

\author{K.~Manukyan}
\affiliation{Department of Physics and Astronomy, University of Notre Dame, Notre Dame, Indiana 46556, USA}

\author{M.~Matney}
\affiliation{Department of Physics and Astronomy, University of Notre Dame, Notre Dame, Indiana 46556, USA}

\author{J.~O'Neill}
\affiliation{School of Physics and Astronomy, University of Birmingham, Birmingham, B15 2TT, United Kingdom}

\author{J.~Rufino}
\affiliation{Department of Physics and Astronomy, University of Notre Dame, Notre Dame, Indiana 46556, USA}

\author{A.~T.~Sanchez}
\affiliation{Department of Physics and Astronomy, University of Notre Dame, Notre Dame, Indiana 46556, USA}

\author{M.~Williams}
\affiliation{School of Mathematics and Physics, University of Surrey, Guildford, GU2 7XH, United Kingdom}

\author{L.~Zimmer}
\affiliation{Department of Physics and Astronomy, University of Notre Dame, Notre Dame, Indiana 46556, USA}

\date{\today}

\pacs{}

\maketitle


\section{Resonance scan fitting}

The resonance scans were fitted using Equation~(1) in the main paper. For the scan fits, the fit parameters corresponding to the resonance energies ($E^{\rm lab}_{\rm R}$) and relative intensities ($h$) were initialized using the values from Columns II and IV of Table~I in~\cite{Dubois1963}, respectively, and physically reasonable ranges were assigned to these parameters during the fitting procedure.

The target thickness in energy units, $\Delta E^{\rm lab}$, was calculated using the effective target thickness of $\approx$71~$\mu$g/cm$^2$ and the corresponding energy loss obtained using \texttt{SRIM-2013} calculations~\cite{ziegler2010srim}.
The calculated value was used as the initial parameter for the fit, and a physically reasonable range for $\Delta E^{\rm lab}$ was assigned during the fitting procedure.

The parameter $\delta_L$ includes contributions from the beam-energy resolution, Doppler broadening, and the intrinsic resonance width ($\Gamma$). The beam-energy resolution is $\approx$0.3~keV, and the upper limit on $\Gamma$ is 0.1~keV (see Subsection~IV.B.4). The Doppler broadening was calculated using~\cite{uhrmacher1985energy},
\begin{equation}
    \sigma_{\rm Doppler}=\sqrt{\frac{2m_pE_pk_BT}{m_t}},
    \label{eq_dopp}
\end{equation}
where $m_p$ and $m_t$ are the projectile and target masses, respectively, $k_B$ is the Boltzmann constant, and $T$ is the target surface temperature. The target temperature was assumed to be approximately 80$^\circ$C, based on previous experience with similar setups and experimental conditions. The resulting Doppler broadening was estimated to be in the range of 25~eV (at $E_p=525$~keV) to 42~eV (at $E_p=1272$~keV). Therefore, $\delta_L$ was constrained to have a lower limit of 0.3~keV during the fitting procedure.
The parameter $\delta_R$ includes the same contributions as $\delta_L$ but is dominated by beam-energy straggling and target non-uniformity; therefore, it was also constrained to have a lower limit of 0.3~keV during the fitting procedure.

For all resonance scans shown in Figs.~4--6 of the main paper, the above fitting procedure was applied. The fit parameters were initialised using the available experimental information and analytical estimates, allowed to vary within physically reasonable ranges, and subsequently constrained based on the best-fit results, as shown in Figs. 4--6.

\section{$E^{\rm{lab}}_{\rm{R}}$ = 917.4 keV resonance scan}

\begin{figure*}
   \centering
    \includegraphics[width=0.85\textwidth]{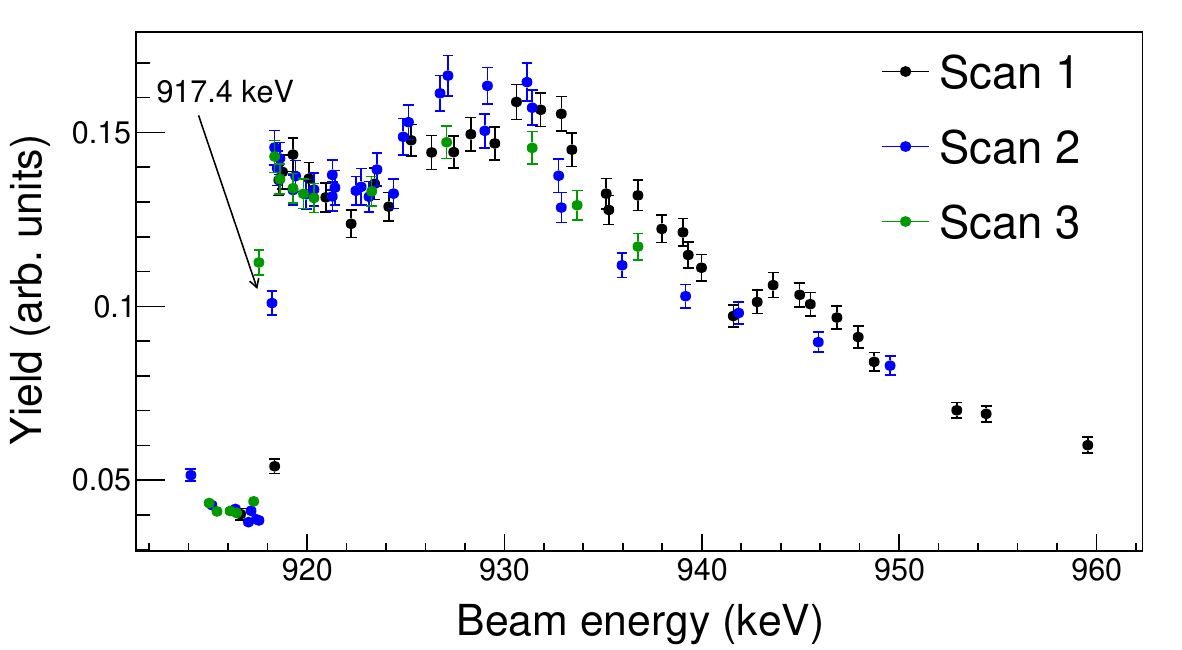}
  \caption{Resonance scan plot for the $E^{\rm{lab}}_{\rm{R}}$~=~917~keV resonance using the 889~keV secondary $\gamma$-ray transition. This resonance was scanned three times during the experiment. 
  }
  \label{fig:Fig-Scan_917keV_Multiple}
\end{figure*}

To check possible target degradation, the 917.4~keV resonance was scanned three times, with each scan requiring $\approx$13--15 hours of beam time on target\footnote{For the scans shown in Figs.~4--6 of the main paper, $\approx$13--15 hours of beam time on target were required, with the exception of the scan shown in Fig.~7, which required $\approx$2 days due to the lower cross section at lower energies.}. The results show no measurable degradation, as shown in Fig.~\ref{fig:Fig-Scan_917keV_Multiple}. The arrow marks the approximate midpoint of the steep rise in yield with increasing beam energy.

\section{Yield calculation}
\begin{figure*}
   \centering
    \includegraphics[width=0.85\textwidth]{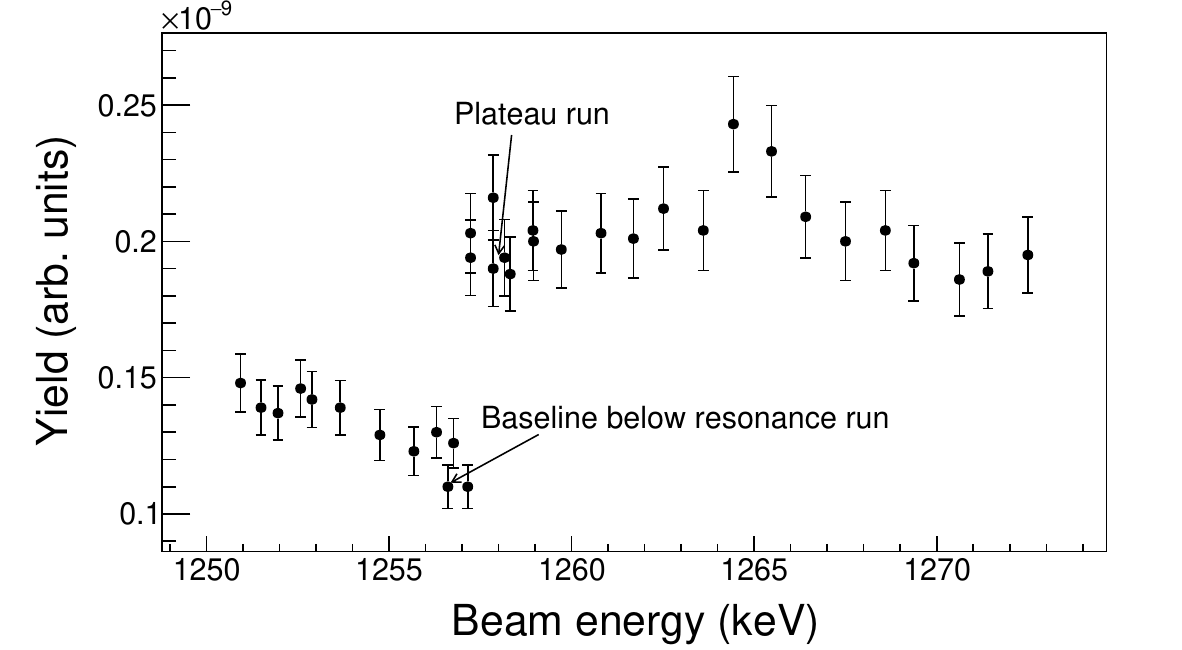}
  \caption{Resonance scan plot for the $E^{\rm{lab}}_{\rm{R}}$~=~1257.5~keV resonance using the 889~keV secondary $\gamma$-ray transition. The plateau run corresponds to a long measurement run performed in far geometry while populating the resonance. The below-resonance baseline run, taken in close geometry, accounts for contributions from nearby resonances that contribute to the background. 
  }
  \label{fig:Fig-Yield}
\end{figure*}

For the calculation of yield, 
\begin{equation}
    Y = Y_{\mathrm{plateau}} - Y_{\mathrm{baseline\,below\,resonance}},
    \label{eq:equation-6}
\end{equation}
where $Y_{\mathrm{plateau}}$ represents the yield on the plateau of the resonance of interest, and $Y_{\mathrm{baseline \, below \, resonance}}$ corresponds to the yield at the baseline preceding the steep rise.
An example for the $E^{\rm{lab}}_{\rm{R}} = 1257.5$ keV resonance is shown in Fig.~\ref{fig:Fig-Yield}.

\section{Thermonuclear reaction rate}

The total thermonuclear reaction rate \(N_A \langle\sigma v\rangle\) in the temperature range 0.1 \(< T_9 <\) 10  calculated using Eq.~(9) in the main paper, is provided in Tables~\ref{tab:Table-1.1} and~\ref{tab:Table-1.2}.

Table~\ref{tab:Table-1.1} contains the reaction rate obtained from the 11 directly measured resonances, corresponding to the blue curve in Fig.~8 of the main paper. Table~\ref{tab:Table-1.2} includes the reaction rate obtained from the same 11 directly measured resonances together with 7 resonances inferred from transfer-reaction data, also shown as the blue curve in Fig.~8 of the main paper.
\begin{table*}
\begin{ruledtabular}
\caption{Total thermonuclear reaction rate $N_A \langle\sigma v\rangle$ for the 
$^{45}$Sc($p,\gamma$)$^{46}$Ti reaction obtained from the 11 directly measured resonances in units of $\rm{cm^3 s^{-1} mol^{-1}}$, as a function of temperature
$T_9$ .} 
\begin{center}
\begin{tabular}{ l l l l }              
$T_9$  &       Low Rate       &     Median Rate      &      High Rate       \\ \hline
0.1   & $3.61 \times 10^{-26}$ & $9.20 \times 10^{-25}$ & $2.15 \times 10^{-23}$ \\
  0.11  & $1.25 \times 10^{-23}$ & $2.45 \times 10^{-22}$ & $4.40 \times 10^{-21}$ \\
  0.12  & $1.62 \times 10^{-21}$ & $2.54 \times 10^{-20}$ & $3.67 \times 10^{-19}$ \\
  0.13  & $9.83 \times 10^{-20}$ & $1.28 \times 10^{-18}$ & $1.53 \times 10^{-17}$ \\
  0.14  & $3.30 \times 10^{-18}$ & $3.66 \times 10^{-17}$ & $3.74 \times 10^{-16}$ \\
  0.15  & $6.88 \times 10^{-17}$ & $6.66 \times 10^{-16}$ & $5.92 \times 10^{-15}$ \\
  0.16  & $9.78 \times 10^{-16}$ & $8.38 \times 10^{-15}$ & $6.60 \times 10^{-14}$ \\
  0.17  & $1.01 \times 10^{-14}$ & $7.81 \times 10^{-14}$ & $5.52 \times 10^{-13}$ \\
  0.18  & $8.09 \times 10^{-14}$ & $5.66 \times 10^{-13}$ & $3.64 \times 10^{-12}$ \\
  0.19  & $5.17 \times 10^{-13}$ & $3.32 \times 10^{-12}$ & $1.96 \times 10^{-11}$ \\
  0.2   & $2.75 \times 10^{-12}$ & $1.63 \times 10^{-11}$ & $8.93 \times 10^{-11}$ \\
  0.3   & $1.10 \times 10^{-7}$  & $3.97 \times 10^{-7}$  & $1.33 \times 10^{-6}$  \\
  0.4   & $2.59 \times 10^{-5}$  & $7.11 \times 10^{-5}$  & $1.84 \times 10^{-4}$  \\
  0.5   & $8.30 \times 10^{-4}$  & $1.89 \times 10^{-3}$  & $4.13 \times 10^{-3}$  \\
  0.6   & $9.83 \times 10^{-3}$  & $1.95 \times 10^{-2}$  & $3.73 \times 10^{-2}$  \\
  0.7   & $6.40 \times 10^{-2}$  & $1.14 \times 10^{-1}$  & $1.97 \times 10^{-1}$  \\
  0.8   & $2.77 \times 10^{-1}$  & $4.55 \times 10^{-1}$  & $7.32 \times 10^{-1}$  \\
  0.9   & $8.97 \times 10^{-1}$  & $1.39 \times 10^{0}$   & $2.11 \times 10^{0}$   \\
  1     & $2.33 \times 10^{0}$   & $3.48 \times 10^{0}$   & $5.05 \times 10^{0}$   \\
  2     & $1.76 \times 10^{2}$   & $2.31 \times 10^{2}$   & $2.92 \times 10^{2}$   \\
  3     & $6.45 \times 10^{2}$   & $8.30 \times 10^{2}$   & $1.02 \times 10^{3}$   \\
  4     & $1.11 \times 10^{3}$   & $1.42 \times 10^{3}$   & $1.74 \times 10^{3}$   \\
  5     & $1.44 \times 10^{3}$   & $1.83 \times 10^{3}$   & $2.23 \times 10^{3}$   \\
  6     & $1.63 \times 10^{3}$   & $2.07 \times 10^{3}$   & $2.52 \times 10^{3}$   \\
  7     & $1.72 \times 10^{3}$   & $2.18 \times 10^{3}$   & $2.65 \times 10^{3}$   \\
  8     & $1.75 \times 10^{3}$   & $2.21 \times 10^{3}$   & $2.69 \times 10^{3}$   \\
  9     & $1.73 \times 10^{3}$   & $2.19 \times 10^{3}$   & $2.66 \times 10^{3}$   \\
  10    & $1.69 \times 10^{3}$   & $2.14 \times 10^{3}$   & $2.59 \times 10^{3}$   \\
\end{tabular}
\label{tab:Table-1.1}
\end{center}
\end{ruledtabular}
\end{table*}

\begin{table*}
\begin{ruledtabular}
\caption{Total thermonuclear reaction rate $N_A \langle\sigma v\rangle$ for the 
$^{45}$Sc($p,\gamma$)$^{46}$Ti reaction obtained from the 11 directly measured resonances together with 7 resonances inferred from transfer-reaction data in units of $\rm{cm^3 s^{-1} mol^{-1}}$, as a function of temperature
$T_9$ .} 
\begin{center}
\begin{tabular}{ l l l l }              
$T_9$ & Lower Rate & Median Rate & High Rate \\ \hline
0.10 & $3.47 \times 10^{-17}$ & $6.10 \times 10^{-15}$ & $2.10 \times 10^{-13}$ \\
0.11 & $6.09 \times 10^{-16}$ & $8.21 \times 10^{-14}$ & $2.18 \times 10^{-12}$ \\
0.12 & $6.73 \times 10^{-15}$ & $7.29 \times 10^{-13}$ & $1.55 \times 10^{-11}$ \\
0.13 & $5.25 \times 10^{-14}$ & $4.72 \times 10^{-12}$ & $8.34 \times 10^{-11}$ \\
0.14 & $3.14 \times 10^{-13}$ & $2.41 \times 10^{-11}$ & $3.63 \times 10^{-10}$ \\
0.15 & $1.53 \times 10^{-12}$ & $1.02 \times 10^{-10}$ & $1.34 \times 10^{-9}$ \\
0.16 & $6.29 \times 10^{-12}$ & $3.72 \times 10^{-10}$ & $4.33 \times 10^{-9}$ \\
0.17 & $2.26 \times 10^{-11}$ & $1.20 \times 10^{-9}$  & $1.26 \times 10^{-8}$ \\
0.18 & $7.22 \times 10^{-11}$ & $3.49 \times 10^{-9}$  & $3.32 \times 10^{-8}$ \\
0.19 & $2.09 \times 10^{-10}$ & $9.27 \times 10^{-9}$  & $8.09 \times 10^{-8}$ \\
0.2 & $5.55 \times 10^{-10}$ & $2.27 \times 10^{-8}$  & $1.84 \times 10^{-7}$ \\
0.3 & $4.73 \times 10^{-7}$  & $9.51 \times 10^{-6}$  & $4.68 \times 10^{-5}$ \\
0.4 & $3.60 \times 10^{-5}$  & $2.71 \times 10^{-4}$  & $9.64 \times 10^{-4}$ \\
0.5 & $9.02 \times 10^{-4}$  & $3.12 \times 10^{-3}$  & $8.29 \times 10^{-3}$ \\
0.6 & $1.01 \times 10^{-2}$  & $2.34 \times 10^{-2}$  & $4.96 \times 10^{-2}$ \\
0.7 & $6.46 \times 10^{-2}$  & $1.23 \times 10^{-1}$  & $2.23 \times 10^{-1}$ \\
0.8 & $2.78 \times 10^{-1}$  & $4.71 \times 10^{-1}$  & $7.75 \times 10^{-1}$ \\
0.9 & $8.99 \times 10^{-1}$  & $1.42 \times 10^{0}$   & $2.18 \times 10^{0}$  \\
1 & $2.34 \times 10^{0}$   & $3.51 \times 10^{0}$   & $5.13 \times 10^{0}$  \\
2 & $1.76 \times 10^{2}$   & $2.31 \times 10^{2}$   & $2.93 \times 10^{2}$  \\
3 & $6.45 \times 10^{2}$   & $8.30 \times 10^{2}$   & $1.02 \times 10^{3}$  \\
4 & $1.11 \times 10^{3}$   & $1.42 \times 10^{3}$   & $1.74 \times 10^{3}$  \\
5 & $1.44 \times 10^{3}$   & $1.83 \times 10^{3}$   & $2.23 \times 10^{3}$  \\
6 & $1.63 \times 10^{3}$   & $2.07 \times 10^{3}$   & $2.52 \times 10^{3}$  \\
7 & $1.72 \times 10^{3}$   & $2.18 \times 10^{3}$   & $2.65 \times 10^{3}$  \\
8 & $1.75 \times 10^{3}$   & $2.21 \times 10^{3}$   & $2.69 \times 10^{3}$  \\
9 & $1.73 \times 10^{3}$   & $2.19 \times 10^{3}$   & $2.66 \times 10^{3}$  \\
10 & $1.69 \times 10^{3}$   & $2.14 \times 10^{3}$   & $2.59 \times 10^{3}$  \\ 
\end{tabular}
\label{tab:Table-1.2}
\end{center}
\end{ruledtabular}
\end{table*}

\bibliographystyle{apsrev4-2}